\documentclass[aps,prd,preprint,nofootinbib,a4paper,11pt,superscriptaddress]{revtex4-1}

\usepackage[T1]{fontenc}
\usepackage[centertags]{amsmath}
\usepackage{amsfonts}
\usepackage{amssymb}
\usepackage[sort&compress]{natbib}% упорядочивает ссылки
\usepackage[pdftex]{hyperref}%hypertex
\usepackage[pdftex]{graphicx}
\usepackage[russian,english]{babel}

\DeclareMathOperator{\re}{Re}
\DeclareMathOperator{\im}{Im}
\DeclareMathOperator{\sgn}{sgn}
\DeclareMathOperator{\Ai}{Ai}

\newcommand{\vf}{\varphi}

\newcommand{\al}{\alpha}
\newcommand{\be}{\beta}
\newcommand{\ga}{\gamma}
\newcommand{\Ga}{\Gamma}
\newcommand{\de}{\delta}

\newcommand{\la}{\lambda}
\newcommand{\La}{\Lambda}
\newcommand{\ups}{\upsilon}

\newcommand{\spk}{\mathbf{k}}
\newcommand{\lan}{\langle}
\newcommand{\ran}{\rangle}

\begin{document}

\selectlanguage{english}

\title{Radiation of de-excited electrons at large times in a strong electromagnetic plane wave}

\date{\today}

\author{P.O. Kazinski}
\email[E-mail:]{kpo@phys.tsu.ru}
\affiliation{Physics Faculty, Tomsk State University, Tomsk, 634050 Russia}
\affiliation{Laboratory of Mathematical Physics, Tomsk Polytechnic University, Tomsk, 634050 Russia}

\begin{abstract}

The late time asymptotics of the physical solutions to the Lorentz-Dirac equation in the electromagnetic external fields of simple configurations -- the constant homogeneous field, the linearly polarized plane wave (in particular, the constant uniform crossed field), and the circularly polarized plane wave -- are found. The solutions to the Landau-Lifshitz equation for the external electromagnetic fields admitting a two-parametric symmetry group, which include as a particular case the above mentioned field configurations, are obtained. General properties of the total radiation power of a charged particle are established. In particular, for a circularly polarized wave and constant uniform crossed fields, the total radiation power in the asymptotic regime is independent of the charge and the external field strength, when expressed in terms of the proper-time, and equals a half of the rest energy of a charged particle divided by its proper-time. The spectral densities of the radiation power formed on the late time asymptotics are derived for a charged particle moving in the external electromagnetic fields of the simple configurations pointed above.

\end{abstract}


\maketitle

\section{Introduction}

The upcoming experimental facilities will allow us to observe experimentally a distinct manifestation of the radiation reaction effects in strong electromagnetic waves \cite{ELI,PiMuHaKermp}. Although a clear-cut signal of the radiation friction was detected already in 1946 \cite{Blewett} from the decreasing of electron's orbit in the betatron, the theoretical and experimental studies of other effects stemming from the radiation reaction are of importance especially for the strong fields where such effects are tangible or even dominate under certain circumstances. One of the consequences of the radiation reaction of charged particles in strong fields is their rapid de-excitation to the state with a minimum radiation. For a constant homogeneous external magnetic field this state is well known (see, e.g., \cite{SokolQED}) and corresponds to a motion of charged particles along the magnetic field lines. In this paper, we continue the investigation of late time asymptotics of the evolution of charged particles started in \cite{lde_sol} (see also \cite{HarGibKer,HarHeiMar,NikishLLsol}). At the large proper-times, a charged particle is de-excited and passes to the asymptotic regime where it emits soft photons. We are about to describe these asymptotic regimes for the electromagnetic fields of simple configurations -- constant homogeneous fields, crossed fields, and a plane wave field -- and to obtain the spectral density of the radiation power for the latter two cases.

As long as in the asymptotic regime the electrons are de-excited and the photons they radiate are chiefly soft (with the energies much lesser than the electron rest energy), the classical theory of radiation should work well \cite{BlochNord,KibbleSoft}. The main instrument that we shall employ to investigate the asymptotics of the evolution of charged particles is the Lorentz-Dirac (LD) equation \cite{Lor,Dir}. In spite of the fact that this equation is ill-famed, we shall show as a byproduct of our study that it can be used to obtain solid predictions about the behaviour of charged particles provided its physical solutions are only taken into account. The notion of a physical solution to the LD equation was introduced in \cite{Bhabha} and later was elaborated by many authors (see, e.g., \cite{Bhabha,Plass,Barut,RohrlBook,GuptaEl,Herrera,Klepik,Endres,Spohn,lde_sol}). The interested reader may consult these references for details, especially Sec. II of \cite{lde_sol} since we shall rely on the results of \cite{lde_sol}. Roughly, the physical solution to the LD equation is such a solution which is an analytic function of the coupling constant (the particle charge) near its zero value. This condition completely rules out runaway solutions to the LD equation. Using the physical solutions to the LD equation, we shall find the asymptotic behaviour of charged particles in the above mentioned field configurations and then, employing these asymptotics, shall obtain the spectral density of radiation.

Apart from the main goal outlined above, we shall also study some general properties of the motion of charged particles in these field configurations with the radiation reaction taken into account. In particular, we shall prove that for such external electromagnetic fields the radiation reaction force tends to diminish the total radiation power such that a charged particle strives for the trajectory with a minimum radiation. It is interesting to note that this property complies with the general principle of the least entropy production for non-equilibrium systems \cite{OnsMach,sd}. In addition, for a constant uniform external electromagnetic field the total radiation power turns out to be a monotonically decreasing function of time. Whereas for a plane electromagnetic wave of circular polarization and for a constant uniform crossed field, the total radiation power expressed in terms of the proper-time is independent of the charge and the external field strength. The only property of a charged particle, which determines this asymptotics, is its mass. Of course, in order to reach this asymptotic regime, the particle must be electrically charged, and the time needed to pass to the asymptotics depends on the charge and the external field strength. The properties of the spectral density of radiation formed of the asymptotics of the physical solutions to the LD equation are also rather curious. It turns out that for constant homogeneous crossed fields the maximum intensity of radiation at a given photon energy does not fall on the orbit plane as one may naively expect for ultrarelativistic charged particles, but is directed at a certain angle to this plane. The explicit expression for this angle will be obtained. As for the radiation of a charged particle in a plane electromagnetic wave of constant amplitude, the intensity of this radiation formed on the late time asymptotics represents a system of rings of maxima and minima alternating each other when projected to the plane orthogonal to the direction of propagation of the wave. The arrangement of these rings depends on the photon energy and so the radiation pattern looks like a circular rainbow. The features of this pattern will be completely described.

Note that the numerical simulations of the radiation produced by a charged particle with the radiation reaction taken into account were performed recently, for example, in \cite{SchlTikh,PiMuHaKermp,PiHaKe} (for the early attempts for the case of a constant homogeneous magnetic field see, e.g, \cite{ShenRad}). In these papers, the so-called Landau-Lifshitz (LL) equation \cite{LandLifshCTF} was used to describe the dynamics of a charged particle. Since these studies were mostly numerical, the late time asymptotics were not vouchsafed to be investigated. In the present paper, we shall give an analytical description of the motion and radiation of charged particles at the large proper-times using the LD equation. This equation is the exact classical effective equation of motion for electrically charged particles and it follows from the energy-momentum conservation law (see, e.g., \cite{Teit}) as opposed to the LL equation violating this conservation law in general. In a recent study \cite{IBMFO}, the nonrelativistic integro-differential equation \cite{FoLeOc} with a special electron form-factor was employed to remedy this deficiency. That is, we can secure the energy-momentum conservation law for the LL equation by introducing the electron form-factor of a special form and, consequently, transforming this differential equation to the integro-differential one. However, this form-factor is unknown in the relativistic case. Remark that the LD equation restricted to the subset of physical solutions also allows numerical simulations as the LL equation (see \cite{Plass,Klepik} and for a recent one \cite{AlcLlE}), although they are more time-consuming.

The paper is organized as follows. In Sec \ref{Notation}, the notation and general formulas regarding the motion of charged particles in classical electrodynamics are given. Then, in Sec. \ref{Asympt}, we shall investigate the asymptotics of the physical solutions to the LD equation for a charged particle moving in constant homogeneous, crossed, or plane wave electromagnetic fields. To provide a comprehensive analysis of the asymptotic behaviour of the evolution, we shall also consider there the solutions to the LL equation for the same field configurations. It is known (see \cite{NikishLLsol,Piazza,HLREKR,HarGibKer,HarHeiMar}) that the LL equation is exactly solvable for such external electromagnetic fields. This will give us a simple mean to find the time needed for a charged particle to reach the asymptotic regime. In Sec. \ref{Asympt_Const}, we shall obtain the late time asymptotics for the motion of a charged particle in a constant homogeneous external electromagnetic field. In Sec. \ref{Asympt_Plane}, the analogous analysis will be given for the field of a plane electromagnetic wave. Besides, in these two sections, we shall find the solutions to the LL equation for the electromagnetic fields admitting a two-parametric symmetry group and generalize slightly the results of \cite{NikishLLsol,Piazza}. Section \ref{Radiation} is devoted to the properties of the total radiation power of a charged particle. Then, in Sec. \ref{Spectr_Dens_Const}, we shall thoroughly analyze the spectral density of the radiation power formed on the asymptotics and the possibilities for its observation in an experiment. After all, in Sec. \ref{Spectr_Dens_Wave}, we shall find the major contribution to the spectral density of radiation of charged particles in the fields of linearly and circularly polarized electromagnetic wave of constant amplitude.

\section{Notation}\label{Notation}

We shall use the following conventions and notation. We choose the system of units such that $c=\hbar=1$. Despite the fact that our treatment is purely classical and the Planck constant will not appear in the formulas, it is convenient to introduce it artificially to make easier a comparison with units used to characterize the experimental installations. The action functional for the charged particle with a charge $e$ and a mass $m$ interacting with the electromagnetic field $A_\mu$ on the Minkowski background $\mathbb{R}^{1,3}$ with the metric $\eta_{\mu\nu}=\mathrm{diag}(1,-1,-1,-1)$ takes the form
\begin{equation}\label{action particl}
    S[x(\tau),A(x)]=-m\int{d\tau\sqrt{\dot{x}^2}}-e\int{d\tau A_\mu
    \dot{x}^\mu}-\frac1{16\pi}\int{d^4xF_{\mu\nu}F^{\mu\nu}},
\end{equation}
where $F_{\mu\nu}:=\partial_{[\mu}A_{\nu]}$ is the strength tensor of the electromagnetic field,
\begin{equation}\label{fmunu}
    F_{\mu\nu}=\left[%
\begin{array}{cccc}
  0 & E_x & E_y & E_z \\
  -E_x & 0 & -H_z & H_y \\
  -E_y & H_z & 0 & -H_x \\
  -E_z & -H_y & H_x & 0 \\
\end{array}%
\right],
\end{equation}
and $x^\mu(\tau)$, $\mu=\overline{0,3}$, is the worldline of a charged particle. In the natural parameterization $\dot{x}^2=1$, the Lorentz-Dirac (LD) equation \cite{Lor,Dir} becomes
\begin{equation}\label{lde_ini}
    m\ddot{x}_\mu=eF_{\mu\nu}\dot{x}^\nu+\frac23e^2(\dddot{x}_\mu+\ddot{x}^2\dot{x}_\mu).
\end{equation}
It is convenient express all the lengths in terms of the Compton wavelength $m^{-1}$ of the charged particle, that is we make a transform
\begin{equation}
    x^\mu\rightarrow m^{-1}x^\mu,\qquad\tau\rightarrow m^{-1}\tau.
\end{equation}
Then the LD equation is rewritten as
\begin{equation}\label{lde}
    \ddot{x}_\mu=\bar{F}_{\mu\nu}\dot{x}^\nu+\la(\dddot{x}_\mu+\ddot{x}^2\dot{x}_\mu),\qquad \bar{F}_{\mu\nu}:=em^{-2}F_{\mu\nu},\quad\la:=2e^2/3,
\end{equation}
where $x^\mu$, $\tau$, $\bar{F}_{\mu\nu}$, and $\la$ are dimensionless quantities. In making estimates, it is useful to have in view that for these dimensionless quantities the lengths are measured in the Compton wavelengths, the field strengths are measured in the units of the Schwinger (critical) field $E_0$, and the energy unit is the rest energy of a particle. For an electron we have
\begin{equation}
\begin{gathered}
    l_C\approx3.86\times 10^{-11}\;\text{cm},\qquad t_C\approx1.29\times 10^{-21}\;\text{s},\qquad m\approx 5.11\times10^5\;\text{eV},\\
    E_0=\frac{m^2}{|e|\hbar}\approx 4.41\times 10^{13}\;\text{G},\qquad
    \omega:=\frac{E}{E_0},\qquad\la=\frac{2\al}3\approx\frac{2}{411},
\end{gathered}
\end{equation}
The up-to-date accelerator facilities are able to accelerate the electrons up to the energies of the order of $50$ GeV. The intensities of the laser fields, which are accessible at the present moment \cite{laser_tod,PiMuHaKermp}, are of the order of  $10^{22}$ W/cm${}^2$ with the photon energies about $1$ eV. These data correspond to
\begin{equation}\label{experiment}
\begin{gathered}
    \ga\approx10^5,\qquad \omega\approx1.47\times10^{-4},\qquad \Omega\approx1.96\times 10^{-6},\qquad\frac{\omega}{\Omega}\approx75.0,\\
    \la\omega\approx7.14\times10^{-7},\qquad\la\omega^2\approx1.05\times10^{-10},
\end{gathered}
\end{equation}
where $\ga$ is the Lorentz factor of the electron and $\Omega$ is the photon energy in the units of the the rest energy of an electron. Henceforth, we omit the bars over the dimensionless electromagnetic fields $\bar{F}_{\mu\nu}$ and all the electromagnetic fields below are $\bar{F}_{\mu\nu}$.

\section{Asymptotics of physical solutions}\label{Asympt}

In this section, we obtain the asymptotics of the physical solutions to the LD equation in the electromagnetic fields admitting a two-parametric symmetry group. Namely, we generalize the results of \cite{lde_sol} to the case of an arbitrary constant homogeneous field and the field of an electromagnetic wave of both linear and circular polarizations. For these field configurations (their concrete form will be specified below) we shall find not only the asymptotics of the physical solutions to the LD equation, but integrate in quadratures the so-called Landau-Lifshitz (LL) equation \cite{LandLifshCTF}, which approximately describes the radiation reaction. The latter property (the integrability of the LL equation) is a consequence of the invariance of the electromagnetic field strength tensor under the two-parametric symmetry group. The LD equation also possesses this symmetry group. However, it is not completely integrable for these fields since it is a differential equation of a higher order as against the LL equation.

\subsection{Constant homogeneous field}\label{Asympt_Const}

Let us consider, at first, a charged particle moving in a constant homogeneous electromagnetic field
\begin{equation}\label{strength_nondeg}
    F^{\mu\nu}=\omega_1e_0^{[\mu} e_1^{\nu]}+\omega_2e_2^{[\mu} e_3^{\nu]},\qquad F^2_{\mu\nu}=\omega_1^2(e^0_\mu e^0_\nu-e^1_\mu e^1_\nu)+\omega_2^2(e^2_\mu e^2_\nu+e^3_\mu e^3_\nu),\qquad (e_\al e_\be)=\eta_{\al\be},
\end{equation}
where $e_\al^\mu$, $\al=\overline{0,3}$, is a tetrad constituted by the eigenvectors of the tensor $(F^2)^\mu_\nu$. The vectors $e^\mu_{0,1}$ correspond to the eigenvalue $\omega^2_1$ of the tensor $(F^2)^\mu_\nu$, while the vectors $e^\mu_{2,3}$ are associated with the eigenvalue $-\omega^2_2$. The eigenvalues are expressed through the Poincar\'{e}-invariants of the electromagnetic field $I_1=\mathbf{E}^2-\mathbf{H}^2$ and $I_2=2(\mathbf{EH})$ in the following way:
\begin{equation}\label{om_1_om_2}
    \omega^2_1=(\sqrt{I_1^2+I_2^2}+I_1)/2,\qquad\omega^2_2=(\sqrt{I_1^2+I_2^2}-I_1)/2.
\end{equation}
The Lorentz transforms leaving the strength tensor \eqref{strength_nondeg} of the electromagnetic field intact form an Abelian group $SO(1,1)\times SO(2)$ with the matrix representation
\begin{equation}
    \La^{\mu\nu}=(e_0^\mu e_0^\nu+e_1^\mu e_1^\nu)\ch\psi+e_0^{(\mu}e_1^{\nu)}\sh\psi+(e_2^\mu e_2^\nu+e_3^\mu e_3^\nu)\cos\vf+e_2^{[\mu}e_3^{\nu]}\sin\vf,
\end{equation}
where $\psi$ and $\vf$ are the group parameters and the round brackets at a pair of indices mean a symmetrization without $1/2$.

It is useful to rewrite the LD equation \eqref{lde} in terms of the variables adjusted to the action of the symmetry group
\begin{equation}\label{subs_eh}
    \mathrm{v}_0=\sqrt{u_e}\ch\psi,\qquad \mathrm{v}_1=\sqrt{u_e}\sh\psi,\qquad \mathrm{v}_2=\sqrt{u_h}\cos\vf,\qquad \mathrm{v}_3=\sqrt{u_h}\sin\vf,
\end{equation}
where $\mathrm{v}_\al:=(e_\al\ups)$ are the projections of the $4$-momentum onto the vectors of the tetrad, $\vf(\tau)$ and $\psi(\tau)$ define a curve on the symmetry group, and
\begin{equation}\label{mass_shell}
   u_e=\mathrm{v}_0^2-\mathrm{v}_1^2,\qquad u_h=\mathrm{v}_2^2+\mathrm{v}_3^2,\qquad u_e(\tau)-u_h(\tau)=1.
\end{equation}
Introducing the notation
\begin{equation}\label{bs}
\begin{gathered}
    \dot{u}_e=:u_e b_e,\qquad \dot{u}_h=:u_h b_h,\\
    \bar{a}_e:=\frac{\dot{\mathrm{v}}_0^2-\dot{\mathrm{v}}_1^2}{u_e},\qquad \bar{a}_h:=\frac{\dot{\mathrm{v}}_2^2+\dot{\mathrm{v}}_3^2}{u_h},
\end{gathered}
\end{equation}
the LD equation can be written in the form of a system of the first order equations (for brevity, we do not denote $\dot{\psi}$ and $\dot{\vf}$ as new variables)
\begin{equation}\label{lde in const}
\begin{gathered}
    \la\ddot{\psi}=\dot{\psi}-\omega_1-\la b_e\dot{\psi},\qquad \la\ddot{\vf}=\dot{\vf}-\omega_2-\la b_h\dot{\vf},\\
    \la\dot{b}_e=b_e-\la b_e^2-2\la u_h(\bar{a}_e-\bar{a}_h),\qquad \la\dot{b}_h=b_h-\la b_h^2-2\la u_e(\bar{a}_e-\bar{a}_h),\\
    \la\dot{\bar{a}}_e=2(\bar{a}_e+\omega_1\dot{\psi})-\la b_e[2\bar{a}_e+(\bar{a}_e-\bar{a}_h)u_h],\qquad \la\dot{\bar{a}}_h=2(\bar{a}_h-\omega_2\dot{\vf})-\la b_h[2\bar{a}_h+(\bar{a}_e-\bar{a}_h)u_e].
\end{gathered}
\end{equation}
The equations in the second line are dependent by virtue of the differential consequence of the mass-shell condition \eqref{mass_shell}
\begin{equation}
    u_eb_e=u_hb_h.
\end{equation}
Notice that Eqs. \eqref{lde in const} are still valid at $\omega_{1,2}=\omega_{1,2}(\tau)$.

The system of equations \eqref{bs} and \eqref{lde in const} possesses a stationary physical solution (at $\omega_{1,2}=const$)
\begin{equation}\label{stat_point}
\begin{gathered}
    u_e=1,\qquad u_h=0,\qquad b_e=0,\qquad b_h=\frac{1-g}{\la},\qquad\bar{a}_e=-\omega_1^2,\qquad\bar{a}_h=-\omega_1^2+\frac{g(g-1)}{2\la^2},\\
    \dot{\psi}=\omega_1,\qquad\dot{\vf}=\frac{\omega_2}{g},
\end{gathered}
\end{equation}
where we have introduced the notation
\begin{equation}
    g:=2^{-1/2}\Big(1+4\la^2\omega_1^2+\sqrt{(1+4\la^2\omega_1^2)^2+16\la^2\omega_2^2}\Big)^{1/2}.
\end{equation}
On substituting to \eqref{subs_eh}, this solution corresponds to a hyperbolic motion along the electric field vector when $\omega_1\neq0$. In the case $\omega_1=0$, the particle moves according to this solution along a straight line in the spacetime. The last formula in \eqref{stat_point} gives the limiting value of the rotation frequency of a charged particle revolving around the magnetic field vector in the system of coordinates associated with the tetrad $e^\mu_a$, where $\mathbf{H}\parallel\mathbf{E}$. The frequency is timed with respect to particle's proper-time. Due to the radiation reaction, this frequency is less than the cyclotron frequency $\omega_2$.

It is not difficult to find the solution to the system of equations \eqref{lde in const} in a small neighbourhood of the point \eqref{stat_point}. It has the form
\begin{equation}\label{asympt_const_gen}
    [\de u_h,\de b_h,\de\bar{a}_e,\de\bar{a}_h,\de\dot{\psi},\de\dot{\vf}]=[u_h(0),\de b_h(0),\de\bar{a}_e(0),\de\bar{a}_h(0),\de\dot{\psi}(0),\de\dot{\vf}(0)]e^{\la^{-1}(1-g)\tau},
\end{equation}
where all the preexponential factors are expressed in terms of $u_h(0)$ as
\begin{equation}\label{asympt_const}
\begin{split}
    \de b_h(0)&=\frac{1-g}{\la}\frac{3g-1+4\la^2\omega_1^2(1+g^{-1})}{5g-4-4\la^2\omega_1^2g^{-1}}\frac{2g-1}{3g-1}u_h(0),\\
    \de\bar{a}_e(0)&=2\omega_1^2(1-g^{-1}) u_h(0),\\
    \de\bar{a}_h(0)&=\frac{g-1}{8\la^2}[g-2-8\la^2\omega_1^2(g^{-1}-g^{-2})]u_h(0)+[3g(g-1)-4\la^2\omega_1^2(3-g^{-1})]\frac{\de b_h(0)}{4\la(2g-1)},\\
    \de\dot{\psi}(0)&=\omega_1(g^{-1}-1)u_h(0),\\
    \de\dot{\vf}(0)&=\frac{\la\omega_2}{g}\frac{\de b_h(0)}{2g-1}.
\end{split}
\end{equation}
The corrections to the ``frequencies'' $\de\dot{\psi}$ and $\de\dot{\vf}$ are of the opposite sign as compared with the main contributions \eqref{stat_point} and tend exponentially to zero. The asymptotics we have found generalizes the known expression \cite{lde_sol} for the electromagnetic field with $I_2=0$ (see \cite{Plass} for the case of a constant homogeneous magnetic field).

Now we consider the LL equation \cite{LandLifshCTF},
\begin{equation}\label{LL}
    \dot{\ups}_\mu=F_{\mu\nu}\ups^\nu+\la(\dot{F}_{\mu\nu}\ups^\nu+F_{\mu\nu}F^{\nu\rho}\ups_\rho-\ups^\la F_{\la\nu}F^{\nu\rho}\ups_\rho\ups_\mu),
\end{equation}
approximating the LD equation, for the same field configuration \eqref{strength_nondeg}. Though the LL equation looks cumbersome in comparison with the LD equation, the LL equation can be completely integrated for a constant homogeneous electromagnetic field \cite{NikishLLsol}. The solutions to the LL equation allow us to describe approximately the dynamics of a charged particle on the whole interval of time. That, in turn, allows us to find the estimates for the time needed for the particle to go to the universal regime \eqref{asympt_const_gen} in terms of the initial data and the values of the external fields.

Making the substitution \eqref{subs_eh} and conducting all the calculations as for the LD equation above, we come to the system of equation of the form \eqref{lde in const} for the LL equation
\begin{equation}\label{LL in const}
    (\psi-\la\omega_1)^{\cdot}=\omega_1,\qquad(\vf-\la\omega_2)^{\cdot}=\omega_2,\qquad
    \dot{u}_e=-2\la(\omega_1^2+\omega_2^2)u_eu_h.
\end{equation}
These equations also hold when $\omega_{1,2}$ are variable. For the constant $\omega_{1,2}$, the system \eqref{LL in const} is easily integrated
\begin{equation}\label{solution_LL const}
\begin{gathered}
    u_e=\left[1-\frac{u_h(0)}{u_e(0)}e^{-2\la(\omega_1^2+\omega_2^2)\tau}\right]^{-1},\qquad u_h=\left[\frac{u_e(0)}{u_h(0)}e^{2\la(\omega_1^2+\omega_2^2)\tau}-1\right]^{-1},\\
    \psi=\psi(0)+\omega_1\tau,\qquad\vf=\vf(0)+\omega_2\tau.
\end{gathered}
\end{equation}
Recall that the quantities $u_e$ and $u_h$ are related by the mass-shell condition \eqref{mass_shell}. The momentum components evolve according to Eq. \eqref{subs_eh}. The time needed for a charged particle to reach the asymptotic regime (the hyperbolic or the helical motion) can be estimated as
\begin{equation}
    \tau\gg[2\la(\omega_1^2+\omega_2^2)]^{-1}\big|\ln\frac{u_h(0)}{u_e(0)}\big|.
\end{equation}
In the leading order in $\la$, the asymptotic regime following from the solutions \eqref{solution_LL const} coincides with the asymptotics \eqref{stat_point}, \eqref{asympt_const} of the physical solutions to the LD equation.

\subsubsection{Other solutions to the LL equation with the same symmetry group}\label{Asympt_Const_Other}

When $\omega_{1,2}$ are the functions of $\tau$, the system of equations \eqref{LL in const} can be integrated in quadratures too. This fact can be employed to generate the exact solutions to the LL equation. Besides, as it was mentioned in \cite{lde_sol}, if one of $\omega_{1,2}$ vanishes and the other one has a special form (see below) then the system \eqref{LL in const} is integrable in quadratures too. Namely, let us assume $\omega_1=0$. Then $\psi=const$ and, making an appropriate Lorentz transform, one can put $\psi=0$. If $\omega_2=\omega_2(x_0)$ in this system of coordinates, i.e., a charged particle moves in a uniform magnetic field, which depends on time and is directed along the $x^1$ axis perpendicular to the trajectory, then the system of equations \eqref{LL in const} is reduced to
\begin{equation}
    (\vf-\la\omega_2)'=\omega_2u_e^{-1/2},\qquad u_e'=-2\la\omega_2^2(u_e-1)u_e^{1/2},\qquad\dot{x}_0=u_e^{1/2},
\end{equation}
where the prime denotes the derivative with respect to $x_0$. These equations are integrable in quadratures and
\begin{equation}
    u_e^{1/2}=\frac{1+u_e^{1/2}(0)-(1-u_e^{1/2}(0))\exp[-2\la\int_0^{x_0}dx\omega_2^2(x)]}{1+u_e^{1/2}(0)+(1-u_e^{1/2}(0))\exp[-2\la\int_0^{x_0}dx\omega_2^2(x)]},
\end{equation}
where we have put $x_0(0)=0$. Analogously, when $\omega_2=0$, we can set $\vf=0$ by an appropriate Lorentz transform. If $\omega_1=\omega_1(x_2)$ in this frame, i.e., a charged particle moves in a constant electric field, which is directed along the $x^1$ axis and depends only on $x^2$, then the system \eqref{LL in const} can be cast into the form
\begin{equation}
    (\psi-\la\omega_1)'=\omega_1u_h^{-1/2},\qquad u_h'=-2\la\omega_1^2(1+u_h)u_h^{1/2},\qquad\dot{x}_2=u_h^{1/2},
\end{equation}
where the prime means the derivative with respect to $x_2$. These equations are also integrable in quadratures and
\begin{equation}
    u_h^{1/2}=\tg[\arctg u_h^{1/2}(0)-\la\int_0^{x_2}dx\omega_1^2(x)],
\end{equation}
where it is assumed that $x_2(0)=0$. However, we should note that the electromagnetic fields considered here are not the exact vacuum solutions to the Maxwell equations.

\subsection{Plane electromagnetic wave}\label{Asympt_Plane}

Another one field configuration, which is invariant with respect to a two-parametric symmetry group, is the plane electromagnetic wave field. Namely, the electromagnetic field strength tensor
\begin{equation}\label{fmunu_ew}
    F^{\mu\nu}=\omega(x_-)e_-^{[\mu}\left[e_1^{\nu]}\cos\vf(x_-)+e_3^{\nu]}\sin\vf(x_-)\right],\qquad F^2_{\mu\nu}=\omega^2(x_-)e^-_\mu e^-_\nu,
\end{equation}
where $\omega(x_-)$ and $\vf(x_-)$ are the arbitrary functions, $x_-:=(e_-x)$, and
\begin{equation}
    (e_ae_b)=\begin{bmatrix}
               0 & 0 & 0 \\
               0 & -1 & 0 \\
               0 & 0 & -1 \\
             \end{bmatrix},\quad a,b=\{-,1,3\},
\end{equation}
is invariant with respect to the transformations generated by the two commuting matrices belonging to the Lorentz group
\begin{equation}\label{sym_ew}
    \La_1^{\mu\nu}=\eta^{\mu\nu}+r_1e_1^{[\mu}e_-^{\nu]}+\frac{r_1^2}2e_-^\mu e_-^\nu,\qquad \La_3^{\mu\nu}=\eta^{\mu\nu}+r_3e_3^{[\mu}e_-^{\nu]}+\frac{r_3^2}2e_-^\mu e_-^\nu,
\end{equation}
where $r_1$ and $r_3$ are the group parameters. These matrices act on the vectors $e_a^\mu$ as follows
\begin{equation}
    \La_{1,3\,\nu}^\mu e_-^\nu=e_-^\mu,\qquad\La_{1\,\nu}^\mu e_3^\nu=e_3^\mu,\qquad\La_{1\,\nu}^\mu e_1^\nu=e_1^\mu+r_1e_-^\mu,\qquad\La_{3\,\nu}^\mu e_1^\nu=e_1^\mu,\qquad\La_{3\,\nu}^\mu e_3^\nu=e_3^\mu+r_3e_-^\mu,
\end{equation}
whence the invariance of the tensor \eqref{fmunu_ew} under these transformations is evident. Note that the case of a linearly polarized wave corresponds to the choice $\vf=0$.

As in the case of a constant homogeneous electromagnetic field, it is useful to introduce the variables adjusted to the action of the symmetry group. Every vector of the Minkowski space can be expressed as a linear combination of the tetrad vectors
\begin{equation}
    x^\mu=\frac12(x_+e_-^\mu+ x_-e_+^\mu)-x_1e_1^\mu-x_3e_3^\mu,\qquad x^2=x_+x_--x_1^2-x_3^2,
\end{equation}
where  $e_+^\mu$ is the isotropic vector orthogonal to $e_1^\mu$ and $e_3^\mu$ and such that $(e_+e_-)=2$. Hence, the action of the symmetry group \eqref{sym_ew} on the velocity $4$-vector can be parameterized as
\begin{equation}\label{subs_iso}
    \mathrm{v}_1=\mathrm{v}_-r_1,\qquad \mathrm{v}_3=\mathrm{v}_-r_3,\qquad \mathrm{v}_+=\mathrm{v}_-^{-1}+\mathrm{v}_-(r_1^2+r_3^2),
\end{equation}
where the functions $r_1(\tau)$ and $r_3(\tau)$ determine a curve on the symmetry group. Using these variables, the LD equation is reduced to
\begin{equation}\label{lde_in_ew2}
\begin{aligned}
    \dot{\mathrm{v}}_-&=\la\left\{\ddot{\mathrm{v}}_--\left[\frac{\dot{\mathrm{v}}_-^2}{\mathrm{v}_-}+\mathrm{v}_-^3(\dot{r}_1^2+\dot{r}_3^2)\right]\right\},&&\qquad& \dot{\mathrm{v}}_-&=-\int_\tau^\infty dse^{(\tau-s)/\la}\left[\frac{\dot{\mathrm{v}}_-^2}{\mathrm{v}_-}+\mathrm{v}_-^3(\dot{r}_1^2+\dot{r}_3^2)\right],\\
    \dot{r}_1&=\la\ddot{r}_1+2\la(\ln\mathrm{v}_-)^{\cdot}\dot{r}_1+\omega\cos\vf,&&\qquad&\dot{r}_1\mathrm{v}_-^2&=\int_\tau^\infty\frac{ds}{\la}e^{(\tau-s)/\la}\mathrm{v}_-^2\omega\cos\vf,\\
    \dot{r}_3&=\la\ddot{r}_3+2\la(\ln\mathrm{v}_-)^{\cdot}\dot{r}_3+\omega\sin\vf,&&\qquad&\dot{r}_3\mathrm{v}_-^2&=\int_\tau^\infty\frac{ds}{\la}e^{(\tau-s)/\la}\mathrm{v}_-^2\omega\sin\vf,\\
    \dot{x}_-&=\mathrm{v}_-,
\end{aligned}
\end{equation}
where all the subintegral functions are taken at the proper-time $s$. The integro-differential equations are obtained in the standard way (see, e.g., \cite{Plass,Barut,RohrlBook}) from their differential counterparts on the left restricting the LD equation to the subset of its physical solutions. Since
\begin{equation}\label{acceler}
    -\ddot{x}^2=\frac{\dot{\mathrm{v}}_-^2}{\mathrm{v}_-^2}+\mathrm{v}_-^2(\dot{r}_1^2+\dot{r}_3^2),
\end{equation}
and $\mathrm{v}_->0$, the integrand of the integro-differential equation in the first line of \eqref{lde_in_ew2} is nonnegative, vanishes only for a rectilinear uniform motion, and tends to zero when $\mathrm{v}_-\rightarrow0$. Consequently, $\mathrm{v}_-$ decreases until either the particle becomes moving along a straight line in the spacetime or $\mathrm{v}_-\rightarrow0$. A rectilinear uniform motion is not a solution to particle's equations of motion in the electromagnetic wave field. Therefore, only the last case survives. The point $\mathrm{v}_-=0$ is an attractor for the physical solutions to the LD equation and so, for such field configurations, a charged particle will eventually move with a velocity close to the speed of light along the direction of propagation of the electromagnetic wave (in our case, along the $x^2$ axis in the system of coordinates associated with the tetrad).

In order to proceed, we perform the another one change of variables
\begin{equation}\label{rho_def}
    \dot{r}_1=:\rho\cos(\psi+\vf),\qquad \dot{r}_3=:\rho\sin(\psi+\vf),\qquad b:=(\ln\mathrm{v}_-)^{\cdot}.
\end{equation}
Then the LD equation becomes
\begin{equation}\label{lde in ew}
\begin{aligned}
    \la\rho\mathrm{v}_-\psi'&=\omega\sin\psi-\la\rho\mathrm{v}_-\vf',\\
    \la\mathrm{v}_-\rho'&=-\omega\cos\psi+(1-2\la b)\rho,&&\qquad&\rho\mathrm{v}_-^2&=\int_\tau^\infty\frac{ds}{\la}e^{(\tau-s)/\la}\mathrm{v}_-^2\omega\cos\psi, \\
    \la\mathrm{v}_-b'&=b+\la\rho^2\mathrm{v}_-^2,&&\qquad& b&=-\int_\tau^\infty\frac{ds}{\la}e^{(\tau-s)/\la}\rho^2\mathrm{v}_-^2, \\
    \mathrm{v}_-'&=b,
\end{aligned}
\end{equation}
where the prime denotes the derivative with respect to $x_-$. Recall that $\vf'$ is a given function of $x_-$. If in the course of evolution $\omega(x_-)$ tends to the constant value $\omega_0$, the above system of the first order differential equations possesses a singular physical point
\begin{equation}\label{phys_sing_point}
    \mathrm{v}_-=0,\qquad b=0,\qquad\psi=\pi n,\qquad\rho=(-1)^n\omega_0,
\end{equation}
where the integer number $n$ is chosen such that $(-1)^n\omega_0>0$. The integro-differential equations \eqref{lde in ew} allow one to derive useful constraints on the parameters of the particle trajectory when the electromagnetic wave has a constant amplitude $\omega_0>0$. In this case, bearing in mind that $\mathrm{v}_-$ is a nonincreasing function, we obtain
\begin{equation}\label{rho_est}
    \rho\leq\mathrm{v}_-^{-2}(\tau)\int_\tau^\infty\frac{ds}{\la}e^{(\tau-s)/\la}\mathrm{v}_-^2(s)\omega_0 \leq \omega_0 \int_\tau^\infty\frac{ds}{\la} e^{(\tau-s)/\la} \leq\omega_0,\qquad0\geq b\geq-\omega_0^2\mathrm{v}_-^2.
\end{equation}
Recollecting the definition of $b$, the last inequality can be integrated
\begin{equation}\label{vm_restr}
    \mathrm{v}_-(0)\geq\mathrm{v}_-\geq\frac{\mathrm{v}_-(0)}{\sqrt{1+2\la\omega_0^2\mathrm{v}_-^2(0)\tau}}.
\end{equation}
Later on we shall see that the second inequality in this formula turns into the equality for the solutions to the LL equation in the field of a circularly polarized plane electromagnetic wave of the constant amplitude $\omega_0$ (see Eq. \eqref{solution_LL ew}).

Now we find the asymptotics of the physical solutions to the LD equation for a particular case of a circularly polarized plane electromagnetic wave of the constant amplitude
\begin{equation}\label{ew}
    \omega=const>0,\qquad\vf=\Omega x_-+\vf_0.
\end{equation}
It is not difficult to find this asymptotics from the system of equations \eqref{lde in ew}. At large times, the physical solution must tend to the physical singular point \eqref{phys_sing_point}. Substituting the series developments of $\mathrm{v}_-$, $b$, $\rho$, and $\psi$ in terms of the inverse powers of $x_-$ (it is clear from \eqref{vm_restr} that $x_-\rightarrow+\infty$ when $\tau\rightarrow+\infty$) with constant coefficients and the ``initial data'' \eqref{phys_sing_point}, we come to
\begin{equation}\label{asympt_ew_circ_pre}
\begin{split}
    \mathrm{v}_-&=\frac{1}{\la\omega^2 x_-}+\frac{k}{x_-^2}+O(x_-^{-3}),\\
    \rho&=\omega-\left(1+\frac{\Omega^2}{4\omega^2}\right)\frac{2}{\omega x_-^2}-\left(1+\frac{\Omega^2}{4\omega^2}\right)\frac{4\la\omega k}{x_-^3}+O(x_-^{-4}),\\
    \psi&=\frac{\Omega}{\omega^2x_-}+\frac{\la\Omega k}{x_-^2}+O(x_-^{-3}).
\end{split}
\end{equation}
The asymptotics of the physical solutions \eqref{asympt_ew_circ_pre} is parameterized by the only one arbitrary constant $k$, as it should be. Indeed, on integrating \eqref{rho_def}, the two additional arbitrary constants appear in the expressions for the $4$-velocities. Integrating these $4$-velocities, we obtain a set of solutions parameterized by the six arbitrary constants, that is \eqref{asympt_ew_circ_pre} describes all the physical solutions in a vicinity of the singular point \eqref{phys_sing_point}. In the asymptotic regime, the dependence of the solutions on the constant $k$ disappears. Remark that the higher terms of the expansion \eqref{asympt_ew_circ_pre} do not vanish at $k=0$.

Consider the behaviour of a charged particle in the asymptotic regime with a more detail. For definiteness, we suppose that the particle strikes the electromagnetic wave at $x_0(0)=x_2(0)=0$. Integrating the asymptotics \eqref{asympt_ew_circ_pre}, one can obtain in the leading order
\begin{equation}\label{asympt_ew_circ}
\begin{gathered}
    x_-=\left(\frac{2\tau}{\la\omega^2}\right)^{1/2}+O(\tau^{-1/2}),\qquad\mathrm{v}_-=(2\la\omega^2\tau)^{-1/2}+O(\tau^{-3/2}),\\
    \mathrm{v}_1=\frac{\omega}{\Omega}\sin\vf+O(x_-^{-1}),\qquad\mathrm{v}_3=-\frac{\omega}{\Omega}\cos\vf+O(x_-^{-1}),\qquad\mathrm{v}_+=\la\omega^2x_-\left(1+\frac{\omega^2}{\Omega^2}\right)+O(1),\\
    x_0=\frac{\sqrt{2\la\omega^2}}{3}\left(1+\frac{\omega^2}{\Omega^2}\right)\tau^{3/2}+O(\tau).
\end{gathered}
\end{equation}
The dependence of the trajectory on the initial velocity of the particle is lost in the limit of large proper-times. As we see, after the lapse of a certain time, the charged particle will move along the direction of propagation of the electromagnetic wave with the velocity close to the speed of light \cite{NikishLLsol}. The transverse momentum components will rotate in a circle and will be cophased with the electric field vector of the electromagnetic wave. The radius of this circle equals $\omega/\Omega$. In the coordinate space, the particle moves along a helix with the axis parallel to the $x^2$ axis. The radius of the ``circle'', which the particle sweeps up in the $(x^1,x^3)$ plane, grows linearly with $x_-$. Inasmuch as the particle crosses the constant phase surfaces of the electromagnetic wave non-uniformly, the frequency of this rotation (the derivative of $\Omega x_-$) declines as $\tau^{-1/2}$ with respect to the proper-time and as $x_0^{-2/3}$ with respect to the laboratory time.

For a linearly polarized wave of the constant amplitude $\omega_0$,
\begin{equation}\label{ew_linear}
    \omega=\omega_0\cos\psi,\qquad\psi:=\Omega x_-+\psi_0,\qquad \vf=0,
\end{equation}
the system of equations \eqref{lde_in_ew2} is written as
\begin{equation}\label{lde_in_ewlin}
    \la(\rho\mathrm{v}_-^2)'=\mathrm{v}_-(\rho-\omega_0\cos\psi),\qquad\la\mathrm{v}_-\mathrm{v}_-''=\mathrm{v}_-'+\la\rho^2\mathrm{v}_-^2,\qquad\dot{r}_3=0,
\end{equation}
where $\rho\equiv\dot{r}_1$. We shall seek for a solution to these equations in the form of the series
\begin{equation}\label{rho_v_expans}
    \rho=\omega_0\cos\psi+\rho_1/x_-+\rho_2/x_-^2+\ldots,\qquad\mathrm{v}_-=u_1/x_-+u_2/x_-^2+\ldots,
\end{equation}
where $\rho_i$ and $u_i$ are the functions of $x_-$ bounded with all their derivatives at $x_-\rightarrow+\infty$. The leading order of the expansion \eqref{rho_v_expans} is determined by \eqref{lde_in_ewlin} at $\la=0$. Substituting the series \eqref{rho_v_expans} into \eqref{lde_in_ewlin} and collecting the terms at the same power of  $x_-$, we obtain for the first and the second equations
\begin{equation}\label{lde_expand}
\begin{split}
    u_1\big[\rho_1+\la\omega_0\Omega u_1\sin\psi-2\la\omega_0 u'_1 \cos\psi\big]/x_-^2+\ldots & =0,\\
    u'_1/x_-+\big[u'_2-(1+\la u''_1) u_1 +\la\omega^2_0 u_1^2\cos^2\psi\big]/x_-^2+\ldots & =0,
\end{split}
\end{equation}
respectively. As long as the functions $\rho_i$ and $u_i$ are bounded with all their derivatives, the terms at the different powers of $x_-$ should vanish independently. The first term of the second equation entails $u_1=const$. The second term of the second equation gives us $u_2$. Moreover, the boundedness of $u_2$ at $x_-\rightarrow+\infty$ can be provided if and only if $u_1=2(\la\omega_0^2)^{-1}$. Then we find $\rho_1$ from the first equation in \eqref{lde_expand}. If we continue this process, we arrive at
\begin{equation}
\begin{split}
    \rho&=\omega_0\cos\psi-\frac{2\Omega\sin\psi}{\omega_0 x_-}-\frac{2k\Omega\omega^4_0\sin\psi+5\omega^2_0\cos3\psi+(8\Omega^2+11\omega^2_0)\cos\psi}{2\omega^3_0x_-^2}+O(x_-^{-3}),\\ \mathrm{v}_-&=\frac{2}{\la\omega^2_0x_-}+\Big(k-\frac{\sin2\psi}{\Omega\omega^2_0}\Big)(\la x_-^2)^{-1}+O(x_-^{-3}),
\end{split}
\end{equation}
where $k$ is an arbitrary constant. Thus we have found the asymptotics of all the physical solutions to Eqs. \eqref{lde_in_ewlin} at $x_-\rightarrow+\infty$. Assuming $x_0(0)=x_2(0)=0$ and integrating the asymptotics above, we have
\begin{equation}\label{asympt_ew_lin}
\begin{gathered}
    x_-=2\Big(\frac{\tau}{\la\omega^2_0}\Big)^{1/2}+O(\tau^{-1/2}),\qquad\mathrm{v}_-=(\la\omega^2_0\tau)^{-1/2}+O(\tau^{-3/2}),\\
    \mathrm{v}_1=\frac{\omega_0}{\Omega}\sin\psi+O(x_-^{-1}),\qquad\mathrm{v}_3=0,\qquad\mathrm{v}_+=\frac{\la\omega^2_0x_-}2\big(1+\frac{\omega^2_0}{\Omega^2}\sin^2\psi\big)+O(1),\\
    x_0=\frac{\sqrt{\la\omega^2_0}}{3}\big(1+\frac{\omega^2_0}{2\Omega^2}\big)\tau^{3/2}+O(\tau).
\end{gathered}
\end{equation}
The asymptotics of the $4$-velocity differs from the analogous asymptotics \eqref{asympt_ew_circ} in the field of the circularly polarized wave only by the replacement $\omega^2_0\rightarrow2\omega^2$ and $\sin^2\psi\rightarrow1/2$ in the corresponding places. In other words, having averaged over the oscillation period of the wave, these asymptotics for the $4$-velocities coincide.

Let us turn to the solution of the LL equation \eqref{LL} for the electromagnetic field \eqref{fmunu_ew} with the arbitrary functions $\omega(x_-)$ and $\vf(x_-)$. The substitution \eqref{subs_iso} reduces this equation to the system
\begin{equation}
    \dot{\mathrm{v}}_-=-\la\omega^2\mathrm{v}_-^3,\qquad(r_1-\la\omega\cos\vf)^{\cdot}=\omega\cos\vf,\qquad (r_3-\la\omega\sin\vf)^{\cdot}=\omega\sin\vf,\qquad\dot{x}_-=\mathrm{v}_-,
\end{equation}
which is integrable in quadratures \cite{Piazza,HLREKR,NikishLLsol,HarGibKer,HarHeiMar} in an evident way. It is convenient to rewrite this system taking $x_-$ as an evolutionary parameter. Then
\begin{equation}
\begin{split}
    \mathrm{v}_-&=\mathrm{v}_-(0)\Big[1+\la\mathrm{v}_-(0)\int_0^{x_-}dx\omega^2(x)\Big]^{-1},\\
    r_1-\la\omega\cos\vf&=r_1(0)-\la\omega(0)\cos\vf(0)+\mathrm{v}_-^{-1}(0)\int_0^{x_-}dx\omega(x)\cos\vf(x)\Big[1+\la\mathrm{v}_-(0)\int_0^{x}dy\omega^2(y)\Big],\\
    r_3-\la\omega\sin\vf&=r_3(0)-\la\omega(0)\sin\vf(0)+\mathrm{v}_-^{-1}(0)\int_0^{x_-}dx\omega(x)\sin\vf(x)\Big[1+\la\mathrm{v}_-(0)\int_0^{x}dy\omega^2(y)\Big],
\end{split}
\end{equation}
where we have put, for definiteness, $x_-(0)=0$. The first equation determines $x_-(\tau)$ by a quadrature.

In particular, for a circularly polarized electromagnetic wave of constant amplitude \eqref{ew}, we obtain
\begin{equation}\label{solution_LL ew}
\begin{split}
    \mathrm{v}_-&=\frac{\mathrm{v}_-(0)}{\sqrt{1+2\la\omega^2\mathrm{v}_-^2(0)\tau}},\qquad x_-=\frac{\sqrt{1+2\la\omega^2\mathrm{v}_-^2(0)\tau}-1}{\la\omega^2\mathrm{v}_-(0)},\\
    r_1&=r_1(0)+\la\omega(1+\frac{\omega^2}{\Omega^2})(\cos\vf-\cos{\vf_0})+\frac{\omega}{\Omega}\Big[\la\omega^2x_-\sin\vf+\frac{\sin\vf-\sin{\vf_0}}{\mathrm{v}_-(0)} \Big],\\
    r_3&=r_3(0)+\la\omega(1+\frac{\omega^2}{\Omega^2})(\sin\vf-\sin{\vf_0})-\frac{\omega}{\Omega}\Big[\la\omega^2x_-\cos\vf+\frac{\cos\vf-\cos{\vf_0}}{\mathrm{v}_-(0)} \Big].
\end{split}
\end{equation}
In the last two expressions, the first term in the square brackets dominates at large proper-times. Now the characteristic times needed for the charged particle to go to the asymptotic regime can be readily estimated. The projection $\mathrm{v}_-$ of the momentum onto the isotropic vector and the corresponding coordinate cease to depend on the initial data at the proper-times
\begin{equation}\label{tau_c}
    \tau\gg[2\la\omega^2\mathrm{v}_-^2(0)]^{-1}.
\end{equation}
Then the first term in the square brackets in the last two expressions of \eqref{solution_LL ew} prevails over all the other terms in these expressions provided that
\begin{equation}\label{estimns}
    \tau\gg\frac{\la\omega^2}{2\Omega^2},\qquad \tau\gg\frac{\la\Omega^2}{2\omega^2},\qquad\tau\gg\frac{\Omega^2}{2\la\omega^4}r^2_{1,3}(0).
\end{equation}
When these inequalities are fulfilled, the asymptotics of the solutions to the LL equation coincides with the leading asymptotics of the physical solutions to the LD equation \eqref{asympt_ew_circ}. Notice that the first inequality in \eqref{estimns} is equivalent to $\Omega x_-\gg1$ and is satisfied with a good accuracy when the particle crossed at least two ``humps'' of the electromagnetic wave.

The analogous calculations for a linearly polarized wave \eqref{ew_linear} give
\begin{equation}
\begin{split}
    \mathrm{v}_-&=\mathrm{v}_-(0)\Big[1+\frac{\la\omega_0^2\mathrm{v}_-(0)}{4\Omega}(2\Omega x_-+\sin2\psi-\sin2\psi_0) \Big]^{-1},\\
    x_-&\Big(1-\frac{\la\omega_0^2\mathrm{v}_-(0)}{4\Omega}\sin2\psi_0\Big)+\frac{\la\omega_0^2\mathrm{v}_-(0)}{4\Omega^2}(\Omega^2x_-^2+\sin^2\psi-\sin^2\psi_0)=\mathrm{v}_-(0)\tau,\qquad r_3=r_3(0),\\
    r_1&=r_1(0)+\la\omega_0\Big(1+\frac{\omega_0^2}{2\Omega^2}\Big)(\cos\psi-\cos\psi_0)+\\
    &+\Big(\frac{\omega_0}{\Omega \mathrm{v}_-(0)}-\frac{\la\omega_0^3}{4\Omega^2}\sin2\psi_0\Big)(\sin\psi-\sin\psi_0)+\frac{\la\omega_0^3}{2\Omega}x_-\sin\psi.
\end{split}
\end{equation}
The conditions that the charged particle reached the asymptotic regime are equivalent to \eqref{tau_c} and \eqref{estimns}. The corresponding asymptotic solution agrees with the asymptotics \eqref{asympt_ew_lin} of the physical solutions to the LD equation.

\section{Radiation power}\label{Radiation}

Let us analyze the behaviour of the radiation power of a charged particle moving along the trajectories derived in the previous section. At first, we obtain the total radiation power in the asymptotic regime for the charged particle evolving in the crossed electromagnetic fields \cite{lde_sol}
\begin{equation}\label{asympt_cross}
    \mathrm{v}_1\approx\sgn\omega\Big(\frac{\tau}{2\la}\Big)^{1/2},\qquad\mathrm{v}_-\approx(2\la\omega^2\tau)^{-1/2},\qquad\mathrm{v}_+\approx\Big(\frac{\omega^2\tau^3}{2\la}\Big)^{1/2}.
\end{equation}
Under the term the crossed fields we always mean the electromagnetic fields of the form \eqref{fmunu_ew} with $\vf=0$ and $\omega=const$. Using the asymptotics above, we derive
\begin{equation}\label{power_cr}
    \mathcal{P}=-\la\ddot{x}^2\approx\la\omega^2\mathrm{v}_-^2\approx\frac1{2\tau}\equiv\frac{mc^2}{2\tau}\approx\frac12\Big(\frac{\omega^2}{50\la x_0^2}\Big)^{1/5}.
\end{equation}
In the penultimate equality, we have restored the velocity of light and reverted to the dimensional quantities. Observe a curious property of the total radiation power. When it is expressed in terms of particle's proper-time, it depends only on the mass of the particle in the asymptotic regime and is independent of the charge and the external field strength. As a function of the laboratory time $x_0$, the total radiation power does not depend on the particle charge and is determined by the particle mass and the external field since $\omega\sim e$ and $\la\sim e^2$. The total radiation power evaluated by using the asymptotics \eqref{asympt_ew_circ} for a circularly polarized plane electromagnetic wave of constant amplitude turns out to be equal to
\begin{equation}\label{power_ew_cir}
    \mathcal{P}=-\la\ddot{x}^2\approx\la\omega^2\mathrm{v}_-^2\approx\frac{1}{2\tau}\approx\frac{(\la\omega^2)^{1/3}}{(6x_0)^{2/3}}\Big(1+\frac{\omega^2}{\Omega^2}\Big)^{2/3}.
\end{equation}
As in the case of the crossed fields, the total radiation power written in terms of the proper-time is independent of the particle charge and the external field strength in the asymptotic regime. Analogously, for a linearly polarized plane electromagnetic wave of constant amplitude, we have from the asymptotics \eqref{asympt_ew_lin},
\begin{equation}\label{power_ew_lin}
    \mathcal{P}=-\la\ddot{x}^2\approx\la\omega^2_0\mathrm{v}_-^2\cos^2\psi\approx\frac{\cos^2\psi}{\tau}\approx\frac{(\la\omega^2_0)^{1/3}}{(3x_0)^{2/3}}\Big(1+\frac{\omega^2_0}{2\Omega^2}\Big)^{2/3}\cos^2\psi.
\end{equation}
On averaging over the oscillation period of the wave, the total radiation power of a charged particle moving in the linearly polarized wave passes into the corresponding expressions \eqref{power_cr} or \eqref{power_ew_cir}, when the expression for the power is written in terms of the proper-time.

Now we establish some general properties of the total radiation power for the motion in the electromagnetic fields of the form \eqref{strength_nondeg} and \eqref{fmunu_ew}. It is well known \cite{BagMar} that the total radiation power in a constant homogeneous electromagnetic field is an integral of motion for the Lorentz equations
\begin{equation}
    \ddot{x}_\mu=F_{\mu\nu}\dot{x}^\nu\;\;\Rightarrow\;\;\dddot{x}_\mu=\dot{F}_{\mu\nu}\dot{x}^\nu+F_{\mu\nu}\ddot{x}^\nu \;\;\Rightarrow\;\;\frac12(\ddot{x}\ddot{x})^{\cdot}=\ddot{x}^\mu\dot{F}_{\mu\nu}\dot{x}^\nu=0.
\end{equation}
Also, using the representation \eqref{subs_iso}, we obtain for the electromagnetic fields of the form \eqref{fmunu_ew}
\begin{equation}
    \ddot{x}^\mu\dot{F}_{\mu\nu}\dot{x}^\nu=-\mathrm{v}_-^2\big[(\omega\cos\vf)^{\cdot}\dot{r}_1+(\omega\sin\vf)^{\cdot}\dot{r}_3\big].
\end{equation}
Hence, taking into account the Lorentz equations (see \eqref{lde_in_ew2} with $\la=0$)
\begin{equation}
    \dot{r}_1=\omega\cos\vf,\qquad \dot{r}_3=\omega\sin\vf,\qquad \dot{\mathrm{v}}_-=0,
\end{equation}
we have
\begin{equation}
    \ddot{x}^\mu\dot{F}_{\mu\nu}\dot{x}^\nu=-\frac12\mathrm{v}_-^2(\omega^2)^{\cdot}=-\frac12(\mathrm{v}_-^2\omega^2)^{\cdot}.
\end{equation}
As we see, the quantity
\begin{equation}
    \ddot{x}^2+\mathrm{v}_-^2\omega^2
\end{equation}
is the integral of motion for the Lorentz equations with the electromagnetic fields of the form \eqref{fmunu_ew}. However, as long as
\begin{equation}\label{power_Lor}
    \ddot{x}^2=-\dot{x}^\mu F^2_{\mu\nu}\dot{x}^\nu=-\mathrm{v}_-^2\omega^2
\end{equation}
for the Lorentz equations, this integral of motion is trivial. On the other hand, if $\omega$ is a constant on particle's trajectory, for example, for a circularly polarized plane electromagnetic wave of constant amplitude, then the total radiation power is the integral of motion of the Lorentz equations (recall that $\mathrm{v}_-=const$). Comparing \eqref{power_Lor} with \eqref{power_cr}, \eqref{power_ew_cir}, and \eqref{power_ew_lin}, remark an interesting property: according to the LD equation, the total radiation power of a charged particle in the asymptotic regime formally coincides with \eqref{power_Lor}, i.e., at large proper-times, the contribution of the LD force to $\ddot{x}^2$ is negligible. This is the reason why the asymptotics of the solutions to the LL equation is the same as the asymptotics of the physical solutions to the LD equation for such field configurations.

When the radiation reaction is ``turned on'', the total radiation power ceases to be an integral of motion for the fields of the form \eqref{strength_nondeg} and \eqref{fmunu_ew}. For example, differentiating the LD equation \eqref{lde} written for constant homogeneous external fields and convolving the result with $\ddot{x}^\mu$, we arrive at
\begin{equation}
    \frac12\frac{d}{d\tau}\big[\ddot{x}^2-\la\frac{d}{d\tau}\ddot{x}^2 \big]=-\la F^2_{LD},\qquad F^\mu_{LD}:=\dddot{x}^\mu+\ddot{x}^2\dot{x}^\mu.
\end{equation}
Integrating the both parts of this equation, we obtain
\begin{equation}
    \ddot{x}^2(s)-\la\frac{d}{d s}\ddot{x}^2(s)=\ddot{x}^2(0)-\la\frac{d}{d\tau}\ddot{x}^2(0)-2\la \int_0^s dtF^2_{LD}(t).
\end{equation}
Consequently, we can write for the physical solutions to the LD equation
\begin{multline}\label{power_lde}
    -\la\frac{d}{ds}(e^{-s/\la}\ddot{x}^2)=\Big[\ddot{x}^2(0)-\la\frac{d}{d\tau}\ddot{x}^2(0)-2\la\int_0^s dtF^2_{LD}(t)\Big]e^{-s/\la}\;\;\Rightarrow\\
    \Rightarrow\;\;\ddot{x}^2(\tau)=\ddot{x}^2(0)-\la\frac{d}{d\tau}\ddot{x}^2(0)-2\int_0^\infty dse^{-s/\la}\int_0^{s+\tau}dtF^2_{LD}(t).
\end{multline}
In order to get the latter expression, we have integrated the first equation over $s$ in the limits $[\tau,+\infty)$. The last term on the right-hand side of the second equation in \eqref{power_lde} is nonnegative at $\tau\geq0$ and is zero only for a hyperbolic motion, when $F^\mu_{LD}=0$. Therefore, the total radiation power of a charged particle moving in a constant electromagnetic field does not grow (this feature was mentioned in \cite{Shen}, but without a rigorous proof) and remains constant only for the hyperbolic motion.

It is instructive to obtain this result in a different way. Introducing for brevity the notation, $f_\mu:=F_{\mu\nu}\dot{x}^\nu$, and convolving the LD equation \eqref{lde} with $F_{\mu\nu}f^\nu$, we find
\begin{equation}
    f^\mu\dot{f}_\mu=\la(f^\mu\ddot{f}_\mu+\ddot{x}^2f^2)\;\Rightarrow\;f^\mu\dot{f}_\mu-\la(f^\mu\dot{f}_\mu)^{\cdot}=\la(\ddot{x}^2f^2-\dot{f}^2).
\end{equation}
Whence it follows for the physical solutions
\begin{equation}
    f^\mu\dot{f}_\mu=\int_\tau^\infty ds e^{(\tau-s)/\la}(\ddot{x}^2f^2-\dot{f}^2)\geq0,
\end{equation}
where the functions entering the integrand are taken at the proper-time $s$. The last inequality is a consequence of the fact that the expression standing in the round brackets is always nonnegative as one can easily check in the momentary comoving frame. Thus the Lorentz force squared is a nondecreasing function of the proper-time. On the other hand, convolving the LD equation \eqref{lde} with the Lorentz force, we get
\begin{equation}
    f^\mu\ddot{x}_\mu-\la(f^\mu\ddot{x}_\mu)^{\cdot}=f^2\;\Rightarrow\;f^\mu\ddot{x}_\mu=\int_0^\infty\frac{dt}{\la}e^{-t/\la}f^2(\tau+t).
\end{equation}
The left-hand side of the latter equation can be rewritten with the help of the LD equation as
\begin{equation}
    f^\mu\ddot{x}_\mu=(\ddot{x}^\mu-\la F_{LD}^\mu)\ddot{x}_\mu=\ddot{x}^2-\frac{\la}{2}(\ddot{x}^2)^{\cdot}.
\end{equation}
Then, using the standard trick, we come to
\begin{equation}
    -\ddot{x}^2=-2\int_0^\infty\frac{ds}{\la}\int_0^\infty\frac{dt}{\la}e^{-(2s+t)/\la}f^2(\tau+s+t)\leq-2f^2(\tau)\int_0^\infty\frac{ds}{\la}\int_0^\infty\frac{dt}{\la}e^{-(2s+t)/\la} =-f^2(\tau),
\end{equation}
for the physical solutions. From the first equality we see again that the total radiation power does not grow with time. The last inequality says that the total radiation power calculated by using the LD equation is less than the same quantity calculated with the help of the Lorentz equation in the same field.

Taking into account the above analysis of the particle motion in a constant electromagnetic field, we deduce that for $\omega_1=0$ (see \eqref{om_1_om_2}) the total radiation power decreases monotonically to zero, while for $\omega_1\neq0$ it declines monotonically to the constant value
\begin{equation}
    \mathcal{P}=-\la\ddot{x}^2=\la\omega_1^2,
\end{equation}
which corresponds to a hyperbolic motion. A charged particle tends to move along the trajectory with a minimum radiation. Curiously, this property of the evolution of charged particles complies with a general principle of the least entropy production for non-equilibrium systems \cite{OnsMach,sd}.

As far as the nonuniform electromagnetic fields of the form \eqref{fmunu_ew} are concerned, it is hard to prove a monotone decrease of the radiation power with time. Apparently, this property does not even hold in general. Nevertheless, the second property that the radiation reaction force reduces the radiation power can be deduced. Differentiating the expression \eqref{acceler} written in terms of $b$ and $\rho$ with respect to the proper-time and employing the LD equation \eqref{lde in ew}, we have
\begin{equation}
    (\ddot{x}^2)^{\cdot}=\frac{2}{\la}(\ddot{x}^2+\rho\mathrm{v}_-^2\omega\cos\psi).
\end{equation}
This yields for the physical solutions
\begin{equation}
    -\ddot{x}^2=2\int_\tau^\infty\frac{ds}{\la}e^{2(s-\tau)/\la}\rho\mathrm{v}_-^2\omega\cos\psi,
\end{equation}
where the subintegral functions depend on the proper-time $s$. When the amplitude of the electromagnetic wave is the constant $\omega_0$, we arrive at the estimate
\begin{equation}\label{power_est}
    -\ddot{x}^2\leq2\omega_0\int_\tau^\infty\frac{ds}{\la}e^{2(s-\tau)/\la}\rho\mathrm{v}_-^2\leq\omega_0^2\mathrm{v}_-^2(\tau)\leq\omega_0^2\mathrm{v}_-^2(0),
\end{equation}
where we have used the inequality \eqref{rho_est} and taken into account that $\mathrm{v}_-$ decreases monotonically (see Eqs. \eqref{lde_in_ew2}). The penultimate inequality in formula \eqref{power_est} says that the total radiation power of a charged particle moving in a plane electromagnetic wave of constant amplitude does not exceed the total radiation power calculated by using the Lorentz equations for the same field configuration. The inequality \eqref{power_est} bounds the total radiation power from above by a monotonically decreasing function proportional to the square of the Lorentz force. In other words, the radiation reaction results in a lessening of the total radiation power as in the case of constant external electromagnetic fields. The same property was pointed out in \cite{NikishLLsol} for a circularly polarized wave propagating along a constant electric field, where it is possible to find the exact stationary solution \cite{ZeldLD} to the LD equation. It is appealing to give a proof of this property for a general external electromagnetic field or to find any counterexample.

\section{Spectral density of radiation}

\subsection{Constant homogeneous crossed fields}\label{Spectr_Dens_Const}

The spectral density of the radiation power of a charged particle moving along the worldline $x^\mu(\tau)$ is given by the standard formula \cite{LandLifshCTF}
\begin{equation}\label{power_dens}
    d\mathcal{E}(\spk)=-e^2j_\mu^*(k)j^\mu(k)\frac{d\spk}{4\pi^2},\qquad k^2=k_+k_--k_1^2-k_3^2=0,
\end{equation}
where
\begin{equation}\label{j_2}
    j_\mu(k):=\int dx_\mu(\tau) e^{-ik_\nu x^\nu(\tau)},\qquad j_\mu^*(k)j^\mu(k)=\re(j_-^*j_+)-|j_1|^2-|j_3|^2.
\end{equation}
Therefore, in order to find the spectral density of radiation, it is sufficient to obtain an expression for the Fourier transform of the current density. In the case at hand, i.e., for a charged particle moving in a crossed electromagnetic field, $j_3=0$ in the system of coordinates associated with the tetrad. On integrating the asymptotics \eqref{asympt_cross}, one derives
\begin{equation}
    k_\mu x^\mu=\sqrt{\frac2\la}\frac{k_-}{10}\Big(|\omega|\tau^{5/2}-\frac{10}{3}\zeta'\tau^{3/2}+5|\zeta|^2\frac{\tau^{1/2}}{|\omega|}\Big),
\end{equation}
where we have introduced a useful notation
\begin{equation}
    \zeta\equiv\zeta'+i\zeta''\equiv|\zeta|e^{i\vf}:=\frac{\sgn(\omega)k_1+i|k_3|}{k_-}.
\end{equation}
The complex number $\zeta$ does not depend on the energy $k_0$ of the radiated photon and defines uniquely the exit angle of this photon.

Let us start with the Fourier transform of the current density component $j_1$. Other components are calculated in a similar way. Substituting \eqref{asympt_cross} to \eqref{j_2} and introducing the new integration variable
\begin{equation}
    y^2:=|\omega|\tau,
\end{equation}
we have
\begin{equation}\label{j_1_cr}
    j_1=\sqrt{\frac{2}{\la}}\frac{\sgn(\omega)}{|\omega|^{3/2}} \int_{0}^\infty dyy^2e^{-iS},\qquad S:=\sqrt{\frac{2}{\la}}\frac{k_-y}{10|\omega|^{3/2}}\big(y^4-\frac{10}3\zeta'y^2+5|\zeta|^2\big).
\end{equation}
The lower integration limit $y=0$ corresponds to that value of the proper-time when the asymptotic expression \eqref{asympt_const} becomes valid with a high accuracy. We shall find below the condition that guarantees an independence (with a good accuracy) of the spectral density of radiation from the concrete choice of this instant of time. It is a consequence of the fact that the main contribution to the integral \eqref{j_1_cr} comes from a small neighbourhood of the saddle point of the function $S(y)$.

\begin{figure}[t]
\centering

\includegraphics*[width=7cm]{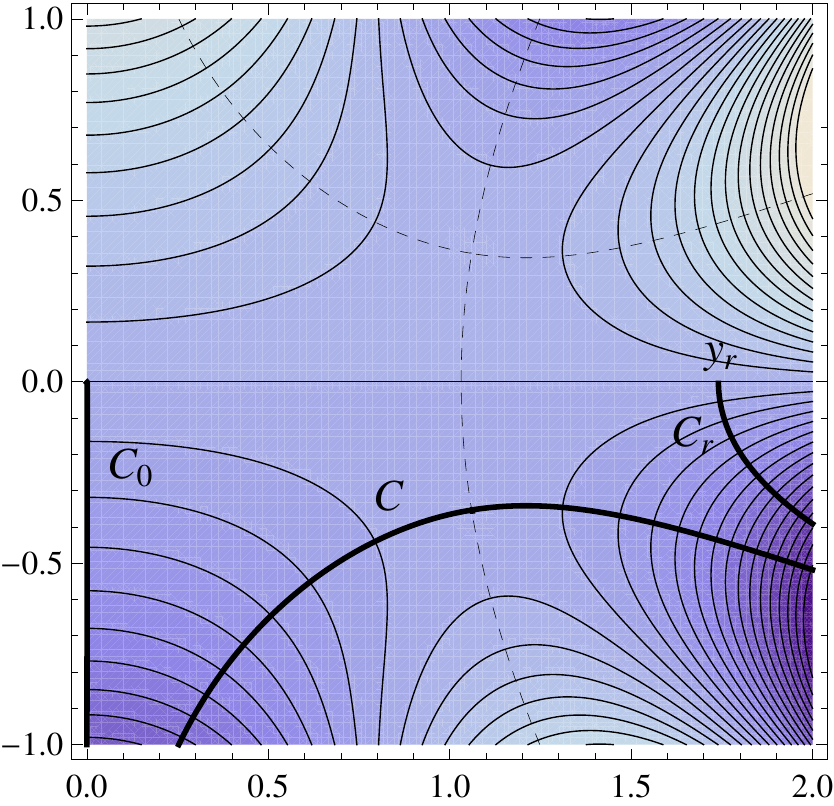}

\caption{{\footnotesize The level lines of the function $\im S$ and the integration contour at $\zeta=1+3i/4$. The initial integration contour $[0,y_r]$ is deformed to the contour $C_0\cup C\cup C_r$. The lines of the steepest descent are depicted as the dashed lines on the figure. One of these lines coincides with the contour $C$. The steepest descent lines intersect at the saddle points.}}
\label{cont_cr}
\end{figure}

The behaviour of the function $S(y)$ in the complex $y$ plane is depicted on Fig. \ref{cont_cr}. Since
\begin{equation}
    S'(y)=\sqrt{\frac{2}{\la}}\frac{k_-}{2|\omega|^{3/2}}(y^2-\zeta)(y^2-\zeta^*),
\end{equation}
the four saddle points arranged symmetrically with respect to the origin of the $y$ plane merge into two points when $\zeta$ is real (i.e., at $k_3=0$). The main contribution to the integral comes from the saddle point $y=\sqrt{\zeta^*}$, this saddle point becoming degenerate when $\zeta$ is real. The optimum integration contour is shown on Fig. \ref{cont_cr}, where we have also introduced the upper integration limit $y_r$ in order to estimate the radiation formation time (see, e.g., \cite{Nikish,Ritus}). A concrete value of the upper integration limit is found from the requirement that, in increasing $y_r$, the contribution to the spectral density of radiation from the contour $[y_r,+\infty)$ becomes negligibly small as compared with the contribution from the contour $[0,y_r]$.

It is useful for a further analysis to distinguish three cases
\begin{equation}\label{cond_i_ii_iii}
    i)\;(20\al)^{2/5}|\zeta'|\gg1,\qquad ii)\;(20\al)^{2/5}|\zeta''|\gg1,\qquad iii)\;(20\al)^{2/5}|\zeta|\ll1,
\end{equation}
where
\begin{equation}
    \al:=\sqrt{\frac{2}{\la}}\frac{k_-}{10|\omega|^{3/2}}.
\end{equation}
These three conditions ought to be considered as the restrictions on the possible observation angles and the energies of photons. In the last case (iii), as we shall show, the integral is saturated in the vicinity of the point $y=0$ and, consequently, it depends severely on the evolution of a charged particle before the moment when this particle passes to the asymptotic regime \eqref{asympt_cross}.

We proceed with the first case. Introduce the new integration variable
\begin{equation}
    t:=(20\al\zeta')^{1/3}(y-\sqrt{\zeta'}).
\end{equation}
Then
\begin{equation}\label{S_cross}
    S=\al\Big[\frac89(\zeta')^{5/2}+5(\zeta'')^2\zeta'^{1/2}\Big]+Bt+\frac{t^3}3+\frac{ht^4}{4}+\frac{h^2t^5}{20},
\end{equation}
where
\begin{equation}
\begin{split}
    B & \equiv bk_3^2  :=5\Big(\frac{\al^2}{20\zeta'}\Big)^{1/3}(\zeta'')^2=\frac{2k_3^2}{\la|\omega|^3(20\al)^{4/3}\zeta'^{1/3}}=\frac{k_3^2}{2|\omega|k_-(\la\sgn(\omega)k_1)^{1/3}},\\
    h & :=(20\al)^{-1/3}(\zeta')^{-5/6}.
\end{split}
\end{equation}
Also we need
\begin{equation}\label{y_2_y_4}
\begin{gathered}
    y^2=\zeta'(1+2ht+h^2t^2),\qquad y^4=\zeta'^2(1+4ht+6h^2t^2+4h^3t^3+h^4t^4),\\
    \int\frac{dt}{2\pi}e^{-i(Bt+t^3/3)}=:\Ai(B),\qquad \int\frac{dt}{2\pi}te^{-i(Bt+t^3/3)}=:i\Ai'(B).
\end{gathered}
\end{equation}
Notice that $h\ll1$ in the case (i). We shall carry out all the calculations taking into account the terms up to the order of $h^2$. It will turn out that only at the order $h^2$ does the first non-vanishing contribution to the spectral density of radiation arise. Substituting \eqref{S_cross} and \eqref{y_2_y_4} to the integral \eqref{j_1_cr} taken on the contour $C$ (see Fig. \ref{cont_cr}) and developing the expression as a series in $h$, we obtain
\begin{equation}
    j_1\approx\sgn(\omega)c\Big[I_0+h\Big(2I_1-\frac{iI_4}{4}\Big)+h^2\Big(I_2-\frac{I_8}{32}-\frac{11i}{20}I_5\Big) \Big],
\end{equation}
where
\begin{equation}
    c:=\frac{(20\al\zeta')^{2/3}}{2k_-}e^{-i\al\big(\frac89(\zeta')^{5/2}+5(\zeta'')^2\zeta'^{1/2}\big)},\qquad I_n:=\int_C dtt^ne^{-i(Bt+t^3/3)}.
\end{equation}
Of course, the contours $C_0$ and $C_r$ also contribute to $j_1$. The magnitude of these contributions will be estimated below, but now we assume that such contributions are small.

Similar calculations for the other current density components
\begin{equation}
    j_-=\sqrt{\frac{2}{\la}}|\omega|^{-3/2} \int_{0}^\infty dye^{-iS},\qquad j_+=\sqrt{\frac{2}{\la}}|\omega|^{-3/2} \int_{0}^\infty dyy^4e^{-iS},
\end{equation}
yield
\begin{equation}
    j_-\approx\frac{c}{\zeta'}\Big[I_0-ih\frac{I_4}{4}-h^2\Big(\frac{I_8}{32}+i\frac{I_5}{20}\Big)\Big],\qquad j_+\approx c\zeta'\Big[I_0+h\Big(4I_1-i\frac{I_4}{4}\Big)+h^2\Big(6I_2-\frac{I_8}{32}-\frac{21i}{20}I_5\Big)\Big].
\end{equation}
Substituting the expressions obtained to \eqref{j_2} and keeping the terms up to the order of $h^2$, we have
\begin{equation}\label{power_dens_cr_m}
\begin{split}
    j_\mu^*(k)j^\mu(k)&\approx4|c|^2|h|^2(\re(I_0^*I_2)-|I_1|^2)=-|4\pi hc|^2(|B||\Ai(B)|^2+|\Ai'(B)|^2),\\
    |hc|^2&=|b|e^{2\al\im[\frac83(\zeta')^{5/2}+5\zeta''^2\zeta'^{1/2}]}=|b|\exp\Big[-\sqrt{\frac{2}\la} \frac{|k_1|^{1/2}}{|k_-\omega|^{3/2}}\Big(\frac8{15}k_1^2+k_3^2\Big)\theta(-\omega k_1) \Big].
\end{split}
\end{equation}
Observe that this expression is symmetric with respect to a sign change of $k_3$, as it should be. Also we see that, at $\sgn(\omega k_1)<0$, this contribution to the spectral density of radiation is exponentially suppressed. This is a consequence of the fact that the most part of the energy of the electromagnetic waves is radiated along the $3$-velocity of a charged particle (the spotlight effect), while the exit direction of the photon and the direction of the $3$-velocity of the particle are always in the different hemispheres for $\sgn(\omega k_1)<0$.

At $\sgn(\omega k_1)\geq0$ (cf. \cite{Ritus,Bord}),
\begin{equation}\label{power_dens_cr}
    j_\mu^*(k)j^\mu(k)\approx-(4\pi)^2b \big\{bk_3^2[\Ai(bk_3^2)]^2+[\Ai'(bk_3^2)]^2\big\},
\end{equation}
where we have shown the dependence on $k_3$ explicitly. The function,
\begin{equation}
    B[\Ai(B)]^2+[\Ai'(B)]^2=\frac{3^{-2/3}}{\Ga^{2}(1/3)}+\frac{3^{2/3}B}{\Ga^{2}(-1/3)}-\frac{\sqrt{3}}{2\pi}B^2+O(B^3)
    =e^{-\frac43B^{3/2}}\Big(\frac{B^{1/2}}{2\pi}+O(1/B)\Big),
\end{equation}
possesses the extremum at the point $B_{ext}\approx0.286$ and
\begin{equation}
    \frac{\{B[\Ai(B)]^2+[\Ai'(B)]^2\}_{B=B_{ext}}}{\{B[\Ai(B)]^2+[\Ai'(B)]^2\}_{B=0}}\approx1.25.
\end{equation}
To put in another way, a charged particle emits a maximum number of photons with a given energy not in the orbit plane (as one may expect for an ultrarelativistic particle), but at the angle specified by the relation
\begin{equation}
    \frac{k_3^2}{k_0^2}=B_{ext}\frac{2|\omega|k_-(\la|k_1|)^{1/3}}{k_0^2}.
\end{equation}
The expression \eqref{power_dens_cr} as a function of the photon energy $k_0$ is monotonically decreasing and singular at $k_0=0$ since $B\sim k_0^{2/3}$ and $b\sim k_0^{-4/3}$. At the small photon energies, the condition (i) in \eqref{cond_i_ii_iii} is violated and the expression \eqref{power_dens_cr} becomes invalid. Transforming the integration measure in the momentum space, the spectral density of radiation can be cast into the form
\begin{equation}\label{power_dens_cr1}
    d\mathcal{E}\approx 4e^2B \big\{B[\Ai(B)]^2+[\Ai'(B)]^2\big\}\frac{k_0^2}{k_3^2}dk_0 d\Omega,
\end{equation}
where $d\Omega$ is the element of a solid angle. The function
\begin{equation}
    B \big\{B[\Ai(B)]^2+[\Ai'(B)]^2\big\}
\end{equation}
possesses the extremum at the point $B'_{ext}\approx0.8$, which determines the energy of photons,
\begin{equation}
    k^0_{ext}=(\la |k_1|/k_0)^{1/2}\frac{(2 B'_{ext} |\omega|k_-/k_0)^{3/2}}{(k_3/k_0)^3},
\end{equation}
where the maximum of radiation occurs at a given exit angle.

Now, let the condition (i) in \eqref{cond_i_ii_iii} be violated, but the inequality (ii) be satisfied. Then the saddle points $y=\sqrt{\zeta}$ and $y=\sqrt{\zeta^*}$ are well separated and the standard WKB method is applicable. In the extremum point $y=\sqrt{\zeta^*}$, we have
\begin{equation}
    S''=-20\al y(y^2-\zeta')=20i\al\zeta''\sqrt{\zeta^*}.
\end{equation}
Using the standard WKB formulas, we derive in the leading order
\begin{equation}
    j_1\approx\Big(\sqrt{\frac{2}{\la}}\frac{\pi\zeta^*}{|k_3||\omega|^{3/2}}\Big)^{1/2}\exp\Big[-i\frac{10}{3}\al\sqrt{\zeta^*}(|\zeta|^2-\frac15\zeta^{*2}) \Big].
\end{equation}
It is easy to verify that the real part of the expression standing in the exponent is negative and so this contribution is exponentially suppressed. The analogous calculations for the rest current density components result in
\begin{equation}
    j_-\approx\frac{j_1}{\zeta^*},\qquad j_+\approx\zeta^*j_1,
\end{equation}
in the leading order. Hence, we obtain
\begin{multline}\label{power_dens_cr_WKB}
    j_\mu^*(k)j^\mu(k)\approx(\re\frac{\zeta^*}{\zeta}-1)|j_1|^2=-2\frac{\zeta''^2}{|\zeta|^2}|j_1|^2
    =-\sqrt{\frac{2}{\la}}\frac{2\pi\sin\vf}{k_-|\omega|^{3/2}}\exp\Big[-\frac{20}{3}\al |\zeta|^{5/2}(\sin\frac{\vf}{2}-\frac15\sin\frac{5\vf}{2})\Big]=\\
    =-\frac{2\pi\sin\vf}{5\al\la|\omega|^3}\exp\Big[-\frac{20}{3}\al |\zeta|^{5/2}(\sin\frac{\vf}{2}-\frac15\sin\frac{5\vf}{2})\Big],
\end{multline}
where, recall, $\vf\in[0,\pi]$ is the phase of $\zeta$.

Consider the case (iii) in \eqref{cond_i_ii_iii}. If this condition is fulfilled then, redefining the integration variable $y\rightarrow\al^{-1/5}y$, it is not difficult to see that the last two terms in the expression standing in the exponent \eqref{j_1_cr} are small. These terms can be taken into account by developing the exponent as a Taylor series in the small parameter and, subsequently, by a termwise integration of the series obtained. The integrals arising in the course of this procedure are reduced to the gamma functions and depend appreciably on the lower integration limit $y=0$. The leading contribution appears to be
\begin{equation}
    j_\mu^*(k)j^\mu(k)\approx\frac{2}{25\la|\omega|^3\al^{6/5}}\Big[\frac{\sqrt{5}-1}{4}\Ga(1/5)-\Ga^2(3/5)\Big].
\end{equation}
However, this expression may change substantially when the contribution of particle's trajectory before the instant $\tau=0$ is taken into account.

Let us estimate the contributions to the spectral density of radiation from the contours $C_0$ and $C_r$. For $C_0$, at
\begin{equation}\label{cond_hv}
    \al^{1/2}|\zeta|\gg1,
\end{equation}
the integral is saturated near the boundary point $y=0$ of the contour. The condition \eqref{cond_hv} is rather close to the first two conditions in \eqref{cond_i_ii_iii}. The contribution from the boundary point is given by the standard expression and in the leading order takes the form
\begin{equation}
    j^{C_0}_1=-\sqrt{\frac{2}{\la}}\frac{\sgn(\omega)}{|\omega|^{3/2}} \int_{-i\infty}^0 dyy^2e^{-iS}\approx i\sgn(\omega)\sqrt{\frac{2}{\la}}\frac{2(5\al|\zeta|^2)^{-3}}{|\omega|^{3/2}}=\sgn(\omega)\frac{8i\la|\omega|^3}{k_-^3|\zeta|^6}.
\end{equation}
In a similar way,
\begin{equation}
\begin{split}
    j^{C_0}_+&=-\sqrt{\frac{2}{\la}}|\omega|^{-3/2} \int_{-i\infty}^0 dyy^4e^{-iS}\approx -i\sqrt{\frac{2}{\la}}\frac{24(5\al|\zeta|^2)^{-5}}{|\omega|^{3/2}}=-\frac{192i\la^2\omega^6}{k_-^5|\zeta|^{10}},\\
    j^{C_0}_-&=-\sqrt{\frac{2}{\la}}|\omega|^{-3/2} \int_{-i\infty}^0 dye^{-iS}\approx -i\sqrt{\frac{2}{\la}}\frac{(5\al|\zeta|^2)^{-1}}{|\omega|^{3/2}}=-\frac{2i}{k_-|\zeta|^{2}}.
\end{split}
\end{equation}
Therefore, the contour $C_0$ makes the contribution to the spectral density of radiation proportional to
\begin{equation}\label{C_0}
    \big[j_\mu^*(k)j^\mu(k)\big]_{C_0}=[\re(j_-^*j_+)-|j_1|^2]_{C_0}\approx\frac{40}{\la|\omega|^3}(5\al|\zeta|^2)^{-6}.
\end{equation}
In the case when this contribution is comparable with or larger than the contributions calculated above for the cases (i) and (ii), the spectral density of radiation depends severely on the evolution of a charged particle before the instant when this particle reaches the asymptotic regime.

For the contour $C_r$ we assume that
\begin{equation}\label{y_r}
    y_r^2=|\omega|\tau_r\gg|\zeta|.
\end{equation}
This means that the point $y_r$ is much more right than the saddle point $y=\sqrt{\zeta^*}$ (see Fig. \ref{cont_cr}) and so the integral is saturated. Then
\begin{equation}
\begin{split}
    j_1^{C_r}&\approx\sqrt{\frac{2}{\la}}\frac{\sgn(\omega)}{|\omega|^{3/2}}\int_{-i\infty}^{y_r}dyy^2e^{-i\al y^5}=-\sqrt{\frac{2}{\la}}\frac{\sgn(\omega)}{|\omega|^{3/2}}\frac{\Ga(3/5,i\al y_r^5)}{5(i\al)^{3/5}},\\
    j_-^{C_r}&\approx\sqrt{\frac{2}{\la}}|\omega|^{-3/2}\int_{-i\infty}^{y_r}dye^{-i\al y^5}=-\sqrt{\frac{2}{\la}}\frac{\Ga(1/5,i\al y_r^5)}{5(i\al)^{1/5}|\omega|^{3/2}},\\
    j_+^{C_r}&\approx\sqrt{\frac{2}{\la}}|\omega|^{-3/2}\int_{-i\infty}^{y_r}dyy^4e^{-i\al y^5}=-\sqrt{\frac{2}{\la}}\frac{\Ga(1,i\al y_r^5)}{5i\al|\omega|^{3/2}}.
\end{split}
\end{equation}
In the cases (i) and (ii) (see \eqref{cond_i_ii_iii}) and when \eqref{y_r} is satisfied, the arguments of the incomplete gamma function are large and so we can employ its asymptotic expansion at large arguments \cite{GrRy},
\begin{multline}\label{C_r}
    \big[j_\mu^*(k)j^\mu(k)\big]_{C_r}=[\re(j_-^*j_+)-|j_1|^2]_{C_r}\approx\\
    \approx\frac{2}{25\la|\omega|^3\al^{6/5}}\Big[-\frac{12}{25}(\al y_r^5)^{-14/5}+O\big((\al y_r^5)^{-24/5}\big)\Big]\approx-\frac{24}{625\la|\omega|^3\al^4y_r^{14}}.
\end{multline}
This contribution is to be compared with the main contribution for the cases (i) and (ii).

In the case (i) at $\sgn(\omega k_1)\geq0$, the main contribution \eqref{power_dens_cr} dominates when (we use for the estimate the value of the spectral density of radiation at $B=0$)
\begin{equation}
    (20\al)^{2/5}|\zeta|\gg\Big(\frac{\pi^2}{3^{2/3} 5120 \Ga^2(1/3)}\Big)^{-3/35}\approx2.15,\qquad (20\al)^{2/5}|\zeta'|\gg 3^{1/4}\Big(\frac{\pi}{8\Ga(1/3)}\Big)^{-3/10}\approx2.34.
\end{equation}
The first inequality is obtained from comparison of \eqref{power_dens_cr} with \eqref{C_0}, while the second inequality follows from comparison of \eqref{power_dens_cr} with \eqref{C_r} and taking into account the inequality \eqref{y_r}. As we see, in the case (i) at $\sgn(\omega k_1)\geq0$, the main contribution to the spectral density of radiation is given by \eqref{power_dens_cr} provided that the radiation has a time to be formed, viz., the inequality \eqref{y_r} holds.

Consider the case (i) at $\sgn(\omega k_1)<0$. To make the estimates we assume that $\zeta''=0$, otherwise at $\zeta''\gg|\zeta'|$ we fall into the case (ii). Keeping this in mind, we deduce that \eqref{power_dens_cr_m} makes the main contribution to the spectral density of radiation when
\begin{equation}
\begin{split}
    \frac{5^{11/14}}{6^{1/7}}\Big(\frac{\pi}{\Ga(1/3)}\Big)^{3/7}\al|\zeta'|^{5/2}e^{-\frac87\al|\zeta'|^{5/2}}&\approx2.94\al|\zeta'|^{5/2}e^{-\frac87\al|\zeta'|^{5/2}}\gg1,\\
    \frac{5}{2^{1/4}3^{5/8}}\Big(\frac{\pi}{\Ga(1/3)}\Big)^{3/4}\al|\zeta'|^{5/2}e^{-2\al|\zeta'|^{5/2}}&\approx2.38\al|\zeta'|^{5/2}e^{-2\al|\zeta'|^{5/2}}\gg1,
\end{split}
\end{equation}
where the first condition follows from comparison of \eqref{power_dens_cr_m} with \eqref{C_0} and the second inequality is obtained from comparison of \eqref{power_dens_cr_m} with \eqref{C_r} taking into account the inequality \eqref{y_r}. The both inequalities cannot be satisfied as long as the expressions entering their left-hand sides are less than unity. Consequently, the expression \eqref{power_dens_cr_m} should not be used for the evaluation of the spectral density of radiation when $\sgn(\omega k_1)<0$.

In the case (ii), comparing \eqref{power_dens_cr_WKB} with \eqref{C_0} and \eqref{C_r}, we see that the WKB answer \eqref{power_dens_cr_WKB} gives the major contribution when
\begin{equation}
    \frac{625}{16384}\frac{\pi}{\sqrt{2|\zeta|}}\big(\al(2|\zeta|)^{5/2}\big)^5e^{-\al(2|\zeta|)^{5/2}}\gg1,\qquad \frac{125}{1536}\frac{\pi}{\sqrt{2|\zeta|}}\big(\al(2|\zeta|)^{5/2}\big)^3e^{-\al(2|\zeta|)^{5/2}}\gg1,
\end{equation}
respectively. To obtain these estimates, we have assumed that $\zeta''\gg|\zeta'|$ and the condition \eqref{y_r} has been used in the second inequality. Notice that the both of these inequalities can be satisfied only if $|\zeta|$ is rather small.

\subsubsection{Experimental verification}\label{Spectr_Dens_Const_Exper}

The contemporary experimental facilities allow one to measure, in principle, the radiation formed on the asymptotics of the physical solutions to the LD equation in the crossed fields. To this aim the intensity of a laser radiation has to be increased by the three orders of magnitude as against one given in \eqref{experiment}. Such intensities will be accessible in the near future \cite{ELI}. To observe the radiation, which we have described above, one should launch the bunch of electrons to the plane electromagnetic wave in such a way that all the time, when the electrons are in the electromagnetic wave, the strength of the electromagnetic field in the vicinity of the electron bunch must change negligibly. This is possible for the ultrarelativistic electrons scattering on the laser photons at an obtuse angle, when the diameter of the laser beam is of the order of a few wavelengths and the characteristic size of the electron bunch is much lesser than the wavelength of the incident photons. Moreover, to conduct a series of experiments with the same strength of the electromagnetic field in the crossing region of the electron and laser beams, these beams have to be synchronized.

Let $N$ be a number of waves that are traversed by the electron moving in the electromagnetic wave, $N\ll1$. Then the proper-time $\tau_{esc}$, that the ultrarelativistic electron spends in the electromagnetic wave, is obtained from the simple system of equations
\begin{equation}\label{tau_esc}
    \ga u\tau_{esc}\sin\al=d,\qquad \ga(1+u\cos\al)\tau_{esc}=\la_\ga N,\qquad\la_\ga:=2\pi\Omega^{-1},
\end{equation}
where $u$ is the module of the electron velocity, $\al$ is the entrance angle of the electron to the laser beam, which is counted from the electromagnetic wave propagation axis, and $d$ is the diameter of the laser beam. All the quantities are taken in the laboratory frame. If
\begin{equation}
    \ga^{-2}\ll\Big( N\frac{\la_\ga}{d}\Big)^{2}\ll u^2,
\end{equation}
then one can put $u=1$ in formulas \eqref{tau_esc} and, in what follows, we assume this is so. The time needed for a charged particle to go to the asymptotic regime is determined by the two characteristic times \cite{lde_sol}
\begin{equation}
    \tau_{1c}:=(2\la\omega^2\mathrm{v}_-(0))^{-1},\qquad\tau_{2c}:=\frac{|\mathrm{v}_1(0)|}{\mathrm{v}_-(0)\omega}.
\end{equation}
The requirement that these characteristic times are much lesser than $\tau_{esc}$ leads to the inequalities
\begin{equation}
    \Big(\frac{d}{\la_\ga}\Big)^2\ll 8\pi N^3\frac{\la\omega^2\ga}{\Omega},\qquad\frac{d}{\la_\ga}\gg\frac{\ga\Omega}{\pi\omega},
\end{equation}
respectively. Taking $\omega=4.65\cdot10^{-3}$, that is greater by one and a half order than in \eqref{experiment}, $\ga=10^3$, and $N=1/5$, we have
\begin{equation}
    \Big(\frac{d}{\la_\ga}\Big)^2\ll 11,\qquad\frac{d}{\la_\ga}\gg0.13,
\end{equation}
and the entrance angle $\al\approx 168.5$ degrees. These inequalities can be satisfied when the diameter of the laser beam is of the order of two wavelengths.

The requirement that the radiation has a time to be formed on the asymptotics, inequality \eqref{y_r} at $\tau_r=\tau_{esc}$, transforms into
\begin{equation}
    |\zeta|\ll\frac{4\pi\omega}{\ga N\Omega}\approx149.
\end{equation}
This results in (reasonable) constraints on the observation angles. The condition (i) in \eqref{cond_i_ii_iii} is convenient to rewrite as
\begin{equation}\label{cond_i_conc}
    k_0\gg\Big[\frac{\la\omega^3(k_-/k_0)^3}{8(k_1/k_0)^5}\Big]^{1/2}\approx\frac{(k_-/k_0)^{3/2}}{(k_1/k_0)^{5/2}}\;4\;\text{eV}.
\end{equation}
As we saw, the expression \eqref{power_dens_cr1} representing the main contribution to the spectral density of radiation formed on the asymptotics possesses the maximum at $B=B'_{ext}$ as a function of the photon energy $k_0$. The maximum spectral density of the radiation formed can be observed when the extremum point lies in the photon energy domain satisfying the inequality \eqref{cond_i_conc}. The condition \eqref{cond_i_conc} restricts the possible values of the parameter $B$ as
\begin{equation}
    \frac{k_3^2}{4k_1^2}\ll B.
\end{equation}
Hence, the maximum of the radiation formed on the asymptotics can be observed if
\begin{equation}
    \frac{k_3^2}{4k_1^2}\ll B'_{ext}.
\end{equation}
This maximum lies in the region of a few dozens of eV,
\begin{equation}
    k^0_{ext}\approx(|k_1|/k_0)^{1/2}\frac{(k_-/k_0)^{3/2}}{(k_3/k_0)^3}\;22.9\;\text{eV},
\end{equation}
and corresponds to soft photons. In this energy region, the radiation is well described by classical theory \cite{BlochNord,KibbleSoft} and, therefore, we may expect that the expressions for the spectral density of radiation obtained above should agree with the one observable in experiments with a high degree of accuracy.

\subsection{Plane electromagnetic wave}\label{Spectr_Dens_Wave}

Let us derive the expressions for the spectral density of radiation formed on the asymptotics of the physical solutions to the LD equation in a plane electromagnetic wave of constant amplitude. We shall not elaborate here a thorough analysis as in the case of the crossed fields (it will be given elsewhere) and shall obtain only the main contribution to the spectral density of radiation formed.

Consider, at first, the radiation of a charged particle in a linearly polarized plane electromagnetic wave. The asymptotics of the physical solutions to the LD equation was found by us in \eqref{asympt_ew_lin}. The corrections to this asymptotics are of the order of $(\Omega x_-)^{-1}$ as compared with the main contribution and negligible provided the electron traverses at least two ``humps'' of the electromagnetic wave. Henceforward, we change the notation $\omega_0\rightarrow\omega$ for brevity. Then, integrating the asymptotics \eqref{asympt_ew_lin} and casting out the terms of the order of $(\Omega x_-)^{-1}$, we arrive at
\begin{equation}
    x_-=2\Big(\frac{\tau}{\la\omega^2}\Big)^{1/2},\qquad x_+=\frac{\la^2\omega^4x_-^3}{12}\big(1+\frac{\omega^2}{2\Omega^2}\big),\qquad x_1=-\frac{\la\omega^3}{2\Omega^2}x_-\cos\psi.
\end{equation}
In general, the analysis of the spectral density of radiation is rather similar to the procedure used in deriving the spectral density of radiation on the solutions to the Lorentz equation in the electromagnetic wave \cite{SokTer,Ritus}. It is useful to change the integration variable in the integrals \eqref{j_2} and introduce the variable $y$ such that
\begin{equation}
    x_-=2y\Big[\la^2\omega^4\big(1+\frac{\omega^2}{2\Omega^2}\big)k_-\Big]^{-1/3}.
\end{equation}
As a result, we have
\begin{equation}\label{B_a_Om}
    k_\mu x^\mu=\frac{y^3}3+y\Big(k_++\frac{\la\omega^3k_1}{\Omega^2}\cos\psi\Big)\Big[\la^2\omega^4\big(1+\frac{\omega^2}{2\Omega^2}\big)k_-\Big]^{-1/3}=:\frac{y^3}3+By+ay\cos\psi
\end{equation}
with
\begin{equation}
    \psi=\tilde{\Omega}y+\psi_0,\qquad\tilde{\Omega}:=2\Omega\Big[\la^2\omega^4\big(1+\frac{\omega^2}{2\Omega^2}\big)k_-\Big]^{-1/3},
\end{equation}
and
\begin{equation}
\begin{split}
    j_- & =2\Big[\la^2\omega^4\big(1+\frac{\omega^2}{2\Omega^2}\big)k_-\Big]^{-1/3}\int_0^\infty dye^{-ik_\mu x^\mu},\\
    j_+ & =\frac{2}{k_-\big(1+\omega^2/2\Omega^2\big)}\int_0^\infty dy y^2\Big(1+\frac{\omega^2}{\Omega^2}\sin^2\psi\Big)e^{-ik_\mu x^\mu},\\
    j_1 & =\frac{2\la\omega^3}{\Omega}\Big[\la^2\omega^4\big(1+\frac{\omega^2}{2\Omega^2}\big)k_-\Big]^{-2/3}\int_0^\infty dy y\sin\psi e^{-ik_\mu x^\mu}.
\end{split}
\end{equation}

We start with the integral defining the Fourier transform of the current density component $j_-$. Using the representation in the form of the Fourier series
\begin{equation}
    e^{-iay\cos\psi}=\sum_{n=-\infty}^\infty\int_{-\pi}^\pi\frac{d\vf}{2\pi}e^{-i(ay\cos\vf-n\psi)}\cos n\vf,
\end{equation}
we can write
\begin{equation}\label{j_-}
    \int_0^\infty dye^{-ik_\mu x^\mu}=\sum_{n=-\infty}^\infty e^{in\psi_0} \int_{-\pi}^\pi d\vf\tilde{\Ai}(B+a\cos\vf-n\tilde{\Omega})\cos n\vf,
\end{equation}
where
\begin{equation}
    \tilde{\Ai}(B):=\int_0^\infty\frac{dt}{2\pi}e^{-i(Bt+t^3/3)}.
\end{equation}
If the argument of this function is positive, the main contribution to the integral defining this function comes from the neighbourhood of the point $t=0$,
\begin{equation}\label{AiryAias}
    \tilde{\Ai}(B)\underset{B\rightarrow+\infty}{\rightarrow}\frac{i}{2\pi B},
\end{equation}
and, consequently, it substantially depends on the form of the evolution before the moment when the charged particle passes to the asymptotic regime. At the large negative argument, the integral is saturated near the saddle point $t=(-B)^{1/2}$,
\begin{equation}\label{AiryAi}
    \tilde{\Ai}(B)\underset{B\rightarrow-\infty}{\rightarrow}\Ai(B)=\big(\pi^2|B|\big)^{-1/4}\sin(2|B|^{3/2}/3+\pi/4)+O(|B|^{-7/4}),
\end{equation}
and weakly depends on the integration limits. We are interested in this last case. From \eqref{B_a_Om} and \eqref{experiment} we see that usually
\begin{equation}
    a\ll B.
\end{equation}
Therefore, in order to observe the radiation formed on the asymptotics, it is necessary to demand
\begin{equation}\label{cond_ewpl}
    1\ll\tilde{\Omega}-B\ll16\pi^2 B^4,
\end{equation}
i.e., the contributions making by the terms with $n\leq0$ in the series \eqref{j_-} are negligibly small in comparison with the contributions from $n\geq1$. Furthermore, already the  term with $n=1$ must depend weakly on the integration limits. The last inequality in \eqref{cond_ewpl} follows from comparison of the term at $n=0$ (see \eqref{AiryAias}) with the term at $n=1$. In the case when the condition \eqref{cond_ewpl} is fulfilled, the Fourier transform of the current density component $j_-$ reduces to
\begin{equation}
    j_-\approx2\Big[\la^2\omega^4\big(1+\frac{\omega^2}{2\Omega^2}\big)k_-\Big]^{-1/3}\sum_{n=1}^\infty e^{in\psi_0} \int_{-\pi}^\pi d\vf\Ai(B+a\cos\vf-n\tilde{\Omega})\cos n\vf.
\end{equation}
Other components of the current density are evaluated in a similar way
\begin{multline}
    j_+\approx\sum_{n=1}^\infty \frac{-2e^{in\psi_0}}{k_-(1+\omega^2/2\Omega^2)}\int_{-\pi}^\pi d\vf\Ai''(B+a\cos\vf-n\tilde{\Omega})\times\hfill\\
    \hfill\times\Big[\big(1+\frac{\omega^2}{2\Omega^2}\big)\cos n\vf-\frac14\cos(n+2)\vf-\frac14\cos(n-2)\vf\Big],\\
    j_1\approx\frac{\la\omega^3}{\Omega}\Big[\la^2\omega^4\big(1+\frac{\omega^2}{2\Omega^2}\big)k_-\Big]^{-2/3}\sum_{n=1}^\infty e^{in\psi_0}\int_{-\pi}^\pi d\vf\Ai'(B+a\cos\vf-n\tilde{\Omega})\times\hfill\\
    \hfill\times\Big[\cos(n-1)\vf-\cos(n+1)\vf\Big].
\end{multline}
Assuming that the different phases $\psi_0$ are equiprobable in the bunch of electrons, we can average over them with a unit weight. Then we have
\begin{multline}\label{power_dens_ewlin}
    \lan\re(j_-^*j_+)\ran\approx-4\la^2\omega^4\Big[\la^2\omega^4\big(1+\frac{\omega^2}{2\Omega^2}\big)k_-\Big]^{-4/3}\sum_{n=1}^\infty \int_{-\pi}^\pi d\vf'\Ai(B+a\cos\vf'-n\tilde{\Omega})\cos n\vf'\times\hfill\\
    \hfill\times\int_{-\pi}^\pi d\vf\Ai''(B+a\cos\vf-n\tilde{\Omega})\Big[\big(1+\frac{\omega^2}{2\Omega^2}\big)\cos n\vf-\frac14\cos(n+2)\vf-\frac14\cos(n-2)\vf\Big],\\
    \lan|j_1|^2\ran\approx \frac{\la^2\omega^6}{\Omega^2}\Big[\la^2\omega^4\big(1+\frac{\omega^2}{2\Omega^2}\big)k_-\Big]^{-4/3}\times\hfill\\
    \hfill\times\sum_{n=1}^\infty\bigg\{\int_{-\pi}^\pi d\vf\Ai'(B+a\cos\vf-n\tilde{\Omega})\Big[\cos(n-1)\vf-\cos(n+1)\vf\Big]\bigg\}^2.
\end{multline}
This, with the account of \eqref{power_dens}, gives the spectral density of radiation formed on the asymptotics of the physical solutions to the LD equation for the field of a linearly polarized plane electromagnetic wave.

The expression obtained can be simplified under the assumption that $a$ is small. Expanding the Airy functions in a Taylor series in $a$, we get the integrals of the form
\begin{equation}
    \int_{-\pi}^\pi d\vf\cos^k\vf\cos n\vf=2^{1-k}\pi C_k^{(k-n)/2},
\end{equation}
at $k\geq n$ and $n+k$ is an even number, otherwise these integrals are equal to zero. Then, retaining only the first correction in $a$ in the expression \eqref{power_dens_ewlin}, we come to
\begin{multline}
    \lan\re(j_-^*j_+)\ran\approx-(2\pi)^2\la^2\omega^4\Big[\la^2\omega^4\big(1+\frac{\omega^2}{2\Omega^2}\big)k_-\Big]^{-4/3}\times\\
    \hfill\times\frac{a^2}{2}\Big\{\Big(\frac{3}{2}+\frac{\omega^2}{\Omega^2}\Big) \Ai'(B-\tilde{\Omega})\Ai'''(B-\tilde{\Omega}) -\frac14[\Ai''(B-2\tilde{\Omega})]^2\Big\},\\
    \lan|j_1|^2\ran\approx(2\pi)^2\frac{\la^2\omega^6}{\Omega^2}\Big[\la^2\omega^4\big(1+\frac{\omega^2}{2\Omega^2}\big)k_-\Big]^{-4/3}\times\hfill\\
    \times\Big\{[\Ai'(B-\tilde{\Omega})]^2+ \Big(\frac{a}2\Big)^2\Big[ \Ai'(B-\tilde{\Omega})\Ai'''(B-\tilde{\Omega}) +[\Ai''(B-2\tilde{\Omega})]^2\Big] \Big\}.
\end{multline}
Taking into account that $\omega/\Omega\gg1$, we eventually arrive at
\begin{multline}\label{power_dens_ewpl}
    \lan\re(j_-^*j_+)-|j_1|^2\ran\approx-\Big(\frac{2\pi}{\omega}\Big)^2\Big(\frac{4\Omega}{\la k_-^2}\Big)^{2/3}\times\\
    \times\Big\{[\Ai'(B-\tilde{\Omega})]^2 + \Big(\frac{a}2\Big)^2\Big[ 3\Ai'(B-\tilde{\Omega})\Ai'''(B-\tilde{\Omega}) +[\Ai''(B-2\tilde{\Omega})]^2\Big] \Big\}.
\end{multline}
The development of the Airy function as a Taylor series in $a$ is rapidly converging provided that
\begin{equation}\label{a_omeg}
    a^2\tilde{\Omega}\ll1.
\end{equation}
This inequality leads to the additional constraint to \eqref{cond_ewpl} on the domain of applicability of the formula \eqref{power_dens_ewpl}. The inequality \eqref{a_omeg} readily follows from the form of the asymptotics of the Airy function \eqref{AiryAi}.

Let us turn to the case of a circularly polarized plane electromagnetic wave of constant amplitude. The analysis is carried out in a compete analogy with the case of a linearly polarized electromagnetic wave. The asymptotics of the physical solutions to the LD equation was obtained by us in \eqref{asympt_ew_circ}. In order to match the notation with the case of a linearly polarized wave, we denote $\vf\rightarrow\psi$. Then, on integrating \eqref{asympt_ew_circ} and discarding the terms of the order of $O((\Omega x_-)^{-1})$ in comparison with the main contribution, we have
\begin{equation}
    x_1\approx-\frac{\la\omega^3}{\Omega^2}x_-\cos\psi,\qquad x_3\approx-\frac{\la\omega^3}{\Omega^2}x_-\sin\psi,\qquad x_+\approx\frac{\la^2\omega^4}3\big(1+\frac{\omega^2}{\Omega^2}\big)x_-^3\qquad\psi=\Omega x_-+\psi_0.
\end{equation}
After the replacement
\begin{equation}
    x_-=y\Big[\frac{\la^2\omega^4}2\big(1+\frac{\omega^2}{\Omega^2}\big)k_-\Big]^{-1/3},
\end{equation}
we obtain
\begin{equation}\label{B_Omega}
    k_\mu x^\mu=\frac{y^3}3+y\Big(\frac{k_+}2+\frac{\la\omega^3k_\perp}{\Omega^2}\cos\tilde{\psi}\Big)\Big[\frac{\la^2\omega^4}2\big(1+\frac{\omega^2}{\Omega^2}\big)k_-\Big]^{-1/3}=:\frac{y^3}3+By+ay\cos\tilde{\psi},
\end{equation}
where
\begin{equation}
    \tilde{\psi}=\tilde{\Omega}y+\psi_0+\bar{\psi}_0,\qquad \cos\bar{\psi}_0=\frac{k_1}{k_\perp},\qquad\tilde{\Omega}:=\Omega\Big[\frac{\la^2\omega^4}2\big(1+\frac{\omega^2}{\Omega^2}\big)k_-\Big]^{-1/3},
\end{equation}
and $k_\perp:=\sqrt{k_1^2+k_3^2}$. The Fourier transforms of the current density components take the form
\begin{equation}
\begin{gathered}
    j_-=\Big[\frac{\la^2\omega^4}2\big(1+\frac{\omega^2}{\Omega^2}\big)k_-\Big]^{-1/3}\int_0^\infty dye^{-ik_\mu x^\mu},\qquad j_+=\frac{2}{k_-}\int_0^\infty dy y^2e^{-ik_\mu x^\mu},\\
    j_1=\frac{\la\omega^3}{\Omega}\Big[\frac{\la^2\omega^4}2\big(1+\frac{\omega^2}{\Omega^2}\big)k_-\Big]^{-2/3}\int_0^\infty dy y\sin\psi e^{-ik_\mu x^\mu},\\ j_3=-\frac{\la\omega^3}{\Omega}\Big[\frac{\la^2\omega^4}2\big(1+\frac{\omega^2}{\Omega^2}\big)k_-\Big]^{-2/3}\int_0^\infty dy y\cos\psi e^{-ik_\mu x^\mu}.
\end{gathered}
\end{equation}
Representing the integrand expressions as Fourier series and assuming that the estimate \eqref{cond_ewpl} is satisfied (for $B$ and $\tilde{\Omega}$ defined in \eqref{B_Omega}), we deduce as in the case of a linearly polarized wave
\begin{multline}\label{jpm}
    \lan\re(j^*_-j_+)\ran\approx-\Big(\frac{2}{k_-}\Big)^{4/3}\Big[\la^2\omega^4\big(1+\frac{\omega^2}{\Omega^2}\big)\Big]^{-1/3}\sum_{n=1}^\infty \int_{-\pi}^\pi d\vf'\Ai(B+a\cos\vf'-n\tilde{\Omega})\cos n\vf'\times\\
    \times\int_{-\pi}^\pi d\vf\Ai''(B+a\cos\vf-n\tilde{\Omega})\cos n\vf.
\end{multline}
We see that the shift of the phase $\psi$ by $\bar{\psi}_0$ is completely canceled out. As for the components $|j_{1,3}|^2$, this is not so:
\begin{multline}\label{j13}
    \lan|j_1|^2\ran\approx\frac{\la^2\omega^6}{4\Omega^2}\Big[\frac{\la^2\omega^4}2\big(1+\frac{\omega^2}{\Omega^2}\big)k_-\Big]^{-4/3}\sum_{n=1}^\infty\bigg|\int_{-\pi}^\pi d\vf \Ai'(B+a\cos\vf-n\tilde{\Omega})\times\hfill\\
    \hfill\times\Big[\cos[(n+1)\vf]e^{i\bar{\psi}_0}-\cos[(n-1)\vf]e^{-i\bar{\psi}_0}\Big]\bigg|^2,\\
    \lan|j_3|^2\ran\approx\frac{\la^2\omega^6}{4\Omega^2}\Big[\frac{\la^2\omega^4}2\big(1+\frac{\omega^2}{\Omega^2}\big)k_-\Big]^{-4/3}\sum_{n=1}^\infty\bigg|\int_{-\pi}^\pi d\vf \Ai'(B+a\cos\vf-n\tilde{\Omega})\times\hfill\\
    \times\Big[\cos[(n+1)\vf]e^{i\bar{\psi}_0}+\cos[(n-1)\vf]e^{-i\bar{\psi}_0}\Big]\bigg|^2.
\end{multline}
Substituting the expressions \eqref{jpm} and \eqref{j13} into \eqref{power_dens}, we obtain the spectral density of radiation formed on the asymptotics of the physical solutions to the LD equation in the field of a circularly polarized plane wave. Notice that this expression is independent of the phase $\bar{\psi}_0$ and represents a function of $k_0$ and $k_\perp$ only. This property is quite expectable from the symmetry arguments for the case of a circularly polarized incident electromagnetic wave.

When $a$ is small and the estimation \eqref{a_omeg} is fulfilled, expanding the expressions obtained in a Taylor series in $a$ and keeping only the leading contribution, we derive
\begin{equation}
\begin{split}
    \lan\re(j^*_-j_+)\ran &\approx -(2\pi)^2\Big(\frac{2}{k_-}\Big)^{4/3}\Big[\la^2\omega^4\big(1+\frac{\omega^2}{\Omega^2}\big)\Big]^{-1/3}\Big(\frac{a}2\Big)^2\Ai'(B-\tilde{\Omega}) \Ai'''(B-\tilde{\Omega}),\\
    \lan |j_1|^2\ran &\approx (2\pi)^2\frac{\la^2\omega^6}{4\Omega^2}\Big[\frac{\la^2\omega^4}2\big(1+\frac{\omega^2}{\Omega^2}\big)k_-\Big]^{-4/3} \bigg\{[\Ai'(B-\tilde{\Omega})]^2+\\
    &+\Big(\frac{a}{2}\Big)^2\big[ (2-\cos2\bar{\psi}_0)\Ai'(B-\tilde{\Omega})\Ai'''(B-\tilde{\Omega})+[\Ai''(B-2\tilde{\Omega})]^2 \big] \bigg\},\\
    \lan |j_3|^2\ran &\approx (2\pi)^2\frac{\la^2\omega^6}{4\Omega^2}\Big[\frac{\la^2\omega^4}2\big(1+\frac{\omega^2}{\Omega^2}\big)k_-\Big]^{-4/3} \bigg\{[\Ai'(B-\tilde{\Omega})]^2+\\
    &+\Big(\frac{a}{2}\Big)^2\big[ (2+\cos2\bar{\psi}_0)\Ai'(B-\tilde{\Omega})\Ai'''(B-\tilde{\Omega})+[\Ai''(B-2\tilde{\Omega})]^2 \big] \bigg\}.
\end{split}
\end{equation}
As a result, taking into account that $\omega/\Omega\gg1$, we arrive at
\begin{multline}\label{power_dens_ew_circ}
    \lan\re(j^*_-j_+)-|j_1|^2-|j_3|^2\ran\approx-\Big(\frac{2\pi}{\omega}\Big)^2\Big(\frac{\sqrt{2}\Omega}{\la k_-^2}\Big)^{2/3}\times\\
    \times\Big\{[\Ai'(B-\tilde{\Omega})]^2 +\Big(\frac{a}{2}\Big)^2\big[4\Ai'(B-\tilde{\Omega})\Ai'''(B-\tilde{\Omega})+[\Ai''(B-2\tilde{\Omega})]^2 \big] \Big\}.
\end{multline}

In the leading order ($a=0$), the spectral density of radiation (formed on the asymptotics) taken at a given fixed photon energy $k_0$ is an oscillating function of $k_\perp$ both for the circularly \eqref{power_dens_ew_circ} and linearly polarized \eqref{power_dens_ewpl} incident electromagnetic waves. The minima the intensity of radiation are specified by the zeroes of the function $\Ai'(x)$:
\begin{equation}
    B-\tilde{\Omega}\approx-1.02; -3.25; -4.82\ldots
\end{equation}
For the fixed energy $k_0$ of photons, the radiation intensity formed on the asymptotics when projected to the plane normal to the direction of propagation of the electromagnetic wave (in our case the $x^2$ axis) will look like the system of concentric rings. In the case of a linearly polarized laser beam, the small corrections proportional to $a$ deform this pattern, but qualitatively it remains the same. Since the maxima of the radiation intensity are observed at the angles which depend on the energy of the emitted photon, the broadband detector will see this radiation as a circular rainbow.

\section{Conclusions}

Concluding the investigation, we may sum up the main results we have achieved. First, we ascertained once again that the LD equation provides a solid basis for the description of the effective dynamics of charged particles. The use of it allowed us to derive the late time asymptotics of particle's motion and to establish some general properties of this motion, which are in accordance with our expectations based on a physical intuition. Nothing pathological happens during the evolution obeying the LD equation, when the physical solutions are only taken into account. Second, we completely described the asymptotic regimes of motion of charged particles in the electromagnetic fields of simple configurations and revealed some peculiar properties of the motion in these regimes. One of these properties is a striking feature of the total radiation power, which, at the large proper-times, turns out to be independent of the charge and the external field strength for a charged particle moving in the electromagnetic plane wave of constant amplitude. Third, employing the asymptotics found, we obtained the spectral density of radiation formed on these asymptotics and established the specific features of such a radiation so that it can be firmly identified in an experiment. This radiation is produced by the de-excited electrons and predominantly constituted by the soft photons. That is why we may expect that the classical theory of radiation is reliable in this regime. Of course, it would be interesting to take the quantum corrections into account or to describe the polarization properties of this radiation, but we leave these topics for a future research.

\begin{acknowledgments}

I appreciate Prof. V.~G. Bagrov for stimulating discussions on the subject. The work is done partially under the project No. 2.3684.2011 of Tomsk State University. It is also supported by the RFBR grant No. 13-02-00551, and by the Russian Ministry of Education and Science, contracts No. 14.B37.21.0911 and No. 14.B37.21.1298.

\end{acknowledgments}

\end{document}